\documentclass[11pt]{article}
\usepackage{jheppub}
\usepackage{graphicx}
\usepackage{amsmath}
\usepackage{amsfonts}
\usepackage{amssymb}
\usepackage{slashed}

\usepackage{hyperref}
\usepackage{bm}

\usepackage[normalem]{ulem}
\usepackage{physics}

\def\qhat{\hat{q}}

\def\xp{x_\perp}
\def\kp{k_\perp}
\def\lp{l_\perp}
\def\qp{q_\perp}
\def\pp{p_\perp}
\def\bkp{\boldsymbol{k}_\perp }

\def\blp{\boldsymbol{l}_\perp }
\def\bpp{\boldsymbol{p}_\perp }
\def\bqp{\boldsymbol{q}_\perp }
\def\bxp{\boldsymbol{x}_\perp }
\def\nbe{n_{\mathrm{B}}}
\def\nfd{n_{\mathrm{F}}}
\def\OO{\mathcal{O}}
\def\als{\alpha_s}
\def\coft{coft}
\def\soft{soft}

\def\qhatmu{\mu}

\def\homu{\rho}

\def\cc{\mathcal{C}}

\def\Bp{B_{\perp}}
\def\bBpt{\boldsymbol{B}_{2\perp}}
\def\bBpo{\boldsymbol{B}_{1\perp}}
\def\bBp{\boldsymbol{B}_{\perp}}
\def\Tr{\mathrm{Tr}}
\def\tmin{\tau_\mathrm{min}}
\def\omt{\omega_T}
\def\qhatz{\qhat_0}
\def\ftime{\tau}
\def\rad{\mathrm{rad}}
\def\tint{\tau_\mathrm{int}}
\def\lmed{L_\mathrm{med}}
\def\Ejet{E}
\def\omc{\omega_c}
\newcommand{\rmi}[1]{{\mbox{\tiny #1}}}
\def\nuuv{\nu_\rmi{UV}}
\def\nuir{\nu_\rmi{IR}}
\def\semi{\mathrm{semi}}
\def\kpir{k^+_\rmi{IR}}
\def\ho{\mathrm{HO}}
\def\subtr{\mathrm{subtr}}
\def\dpp{\frac{d^2 p_{\perp}}{(2\pi)^2}}

\def\dlp{\frac{d^2 l_{\perp}}{(2\pi)^2}}

\def\bq{\boldsymbol{q}}

\title{Classical vs quantum corrections to 
jet broadening in a weakly-coupled 
Quark-Gluon Plasma}
\author[a]{Jacopo Ghiglieri,}
\author[a]{Eamonn Weitz}

\affiliation[a]{
SUBATECH, Universit\'e de Nantes, IMT Atlantique, IN2P3/CNRS,\\
4 rue Alfred Kastler,
La Chantrerie BP 20722, 44307 Nantes, France}

\emailAdd{jacopo.ghiglieri@subatech.in2p3.fr}
\emailAdd{eamonn.weitz@subatech.in2p3.fr}

\abstract{
The transverse momentum
broadening coefficient $\qhat$ receives both soft, classical and radiative, quantum
 corrections. The former are responsible for a large $\OO(g)$ correction, whereas
 the latter enter at relative order
$\als$, but are enhanced by a double logarithm of the length
of the medium over the thermal wavelength. We analyze radiative corrections
for a weakly-coupled quark-gluon plasma. We find that  a thermal  
population of dynamical gluons changes the boundaries and reduces the size of the 
double-logarithmic phase space. It also provides new subdominant 
logarithmic corrections.
We also show how the quantum, double-logarithmic and classical, soft phase spaces
are smoothly connected once the radiated gluon becomes soft enough. Finally,
we discuss a pathway to a determination of radiative corrections
beyond the harmonic-oscillator approximation.
}

\begin{document}
\maketitle
\section{Introduction}
\label{sec:intro}

The highly-energetic, collimated partons that seed 
a jet undergo \textit{transverse momentum broadening} as they propagate through 
the hot QCD medium that is understood to be formed in heavy-ion collisions. 
This broadening is described by the \textit{broadening rate}, or the associated probability
and the \textit{transverse momentum broadening coefficient} $\qhat\equiv \langle k_\perp^2\rangle/L$, which 
describes how much transverse momentum $k_\perp$ is picked up per length $L$ by a 
highly energetic parton propagating through a plasma. The precise characterization of these
quantities is thus  at the forefront
of theoretical and experimental investigation of this medium. We refer to 
\cite{Connors:2017ptx,Cao:2020wlm,Cunqueiro:2021wls,Apolinario:2022vzg} for recent reviews
and 
to~\cite{Burke:2013yra,Andres:2016evc,Xie:2019oxg,Huss:2020whe,Jetscape:2021ehl,Han:2022zxn,Xie:2022ght} for
extractions of $\qhat$ from data.

In a weakly-coupled quark-gluon plasma (QGP), $\qhat$ 
is dominated by Coulomb elastic scatterings with light
partonic medium constituents, $\qhat=\int d^2\kp\,\kp^2\,d\Gamma_\mathrm{el}/d^2\kp$. 
For $\kp\sim T$ the  elastic, differential scattering rate $d\Gamma_\mathrm{el}/d^2\kp$ is of order $g^4T^3/\kp^4$, with $T^3$
the (parametric) light parton density multiplying the Coulomb cross section. The
resulting logarithmic integral is cut off in the infrared (IR) by 
dynamical screening, encoded in Hard Thermal Loop (HTL) resummation
\cite{Braaten:1989mz} at the \emph{soft scale} $gT$. Ultraviolet (UV) regularization is also needed,
as we shall explain later. The soft ($\kp\sim gT$) and \emph{thermal} ($\kp\sim T$) contributions
 were obtained in \cite{Aurenche:2002pd} and \cite{Arnold:2008vd} respectively, 
 and they combine to give $\qhat\sim g^4T^3$ (up to logarithms). 

This leading-order (LO) contribution from the soft scale is the first signal of potentially
large corrections arising from this region. Soft gluons with frequency 
$\omega\ll T$ are distributed on the infrared  tail of the Bose--Einstein
distribution, $\nbe(\omega\ll T)\approx T/\omega\gg 1$. This has two consequences:
for soft, $\omega\sim gT$ modes,  $T/\omega \sim 1/g$ changes the standard 
$g^2$ loop expansion into a $g$ expansion. For \textit{ultrasoft (US) modes}, $\omega \sim g^2T$,
the perturbative expansion breaks down \cite{Linde:1980ts}, but their contribution only enters 
$\qhat$ at relative order $g^2$ \cite{Laine:2012ht}. The second consequence is that this large occupation
number implies the \emph{classical} nature of these contributions. 

In a pioneering study, Caron-Huot showed in \cite{CaronHuot:2008ni} that these classical
plasma effects can be mapped, for the observable at hand, to  three-dimensional,
Euclidean physics described by Electrostatic QCD 
(EQCD)~\cite{Braaten:1994na,Braaten:1995cm,Braaten:1995jr,Kajantie:1995dw,Kajantie:1997tt}.
This paved the way first to the perturbative determination of the $\OO(g)$ next-to-leading 
order (NLO) correction
to $\qhat$,
presented in \cite{CaronHuot:2008ni} itself, without recurring to brute-force numerical
computations in the HTL theory. Second, it was used for non-perturbative determinations
of the transverse scattering rate using lattice EQCD \cite{Panero:2013pla,Moore:2019lgw}
--- see also \cite{Laine:2013lia} --- thus incorporating to all orders
the contribution of classical soft and ultrasoft modes. The impact of this non-perturbative
rate on \emph{medium-induced emission} was examined in \cite{Moore:2021jwe,Schlichting:2021idr}
and found to be very relevant.

Over the past decade, another source of potentially large higher-order corrections 
to $\qhat$ has been discovered \cite{Wu:2011kc,Liou:2013qya,Blaizot:2013vha} --- see \cite{Blaizot:2015lma}
for a review. These 
are \emph{quantum}, \emph{radiative} corrections, arising from keeping track of the recoil
during the medium-induced emission of a gluon. These corrections, while
suppressed by a factor of $g^2$, feature a \emph{double-logarithmic}
enhancement, whose argument was found to be $\lmed/\tmin$, with $\lmed$ the length of the 
medium and  $\tmin\sim 1/T$
for a thermal medium. This potentially large double logarithm can then be resummed,
effectively renormalizing the leading-order value of $\qhat$. 
The evolution equations for the logarithmic resummation 
were analyzed in \cite{Iancu:2014kga,Iancu:2014sha} and 
recently solved numerically in~\cite{Caucal:2021lgf,Caucal:2022fhc},
whereas the change in the argument of the double logarithm from a dense
to a dilute medium was studied in \cite{Blaizot:2019muz}.
\cite{Blaizot:2014bha,Wu:2014nca,Iancu:2014kga} showed that when two collinear gluons are emitted 
with overlapping formation times the same physics is responsible
for double-logarithmic corrections, which are thus universal. This same problem is being
 analyzed in full 
detail, beyond the double-logarithmic terms, 
in \cite{Arnold:2015qya,Arnold:2016kek,Arnold:2016mth,Arnold:2018yjd,Arnold:2018fjr,Arnold:2019qqc,
Arnold:2020uzm,Arnold:2021pin,Arnold:2022epx}.
\cite{Arnold:2021mow} found
that the single-logarithmic corrections too are related
by the same  universality to those determined for $\qhat$ in \cite{Liou:2013qya}.

The presence of two such large corrections, the classical ones at order $g$ and beyond, and the
quantum radiative ones at order $g^2\ln^2(\lmed T)$, together with the 
recent developments on both sides, such as the present availability of high-quality 
non-perturbative data for the classical corrections, naturally begs the question 
of which of the two corrections, if any, can be considered parametrically larger than the other.

Motivated by this question, in this paper we address the connection of the two corrections
in the case of a weakly-coupled quark-gluon plasma. 
We will revisit the calculation 
of radiative corrections with full accounting of the thermal length and momentum scales.
We shall show how the effect of Bose enhancement for the radiated gluon, as well
as the possibility of absorption of thermal gluons
 --- neither was included in previous analyses --- modifies the
shape of the double-log-enhanced region in the frequency--formation time plane, thus
changing the argument of the double logarithm. 
On a more technical standpoint, we also show how the regions giving rise
to classical and quantum corrections meet at their shared boundary, and how 
this can be rephrased in terms of the mapping to the Euclidean 3D theory
introduced by Caron-Huot.

Our analysis will be mostly limited to the largest double-logarithmic corrections only
and it will adopt the so-called harmonic-oscillator approximation:
we shall discuss some smaller corrections and provide a pathway to the 
determinations of missing ones.

The paper is organized as follows: in Sec.~\ref{sec:dlog} we review 
the determination of double-log-enhanced radiative corrections. 
In Sec.~\ref{sec:temperature} we show how this derivation needs 
to be modified to account for the presence of a weakly-coupled dynamical
medium and we present there our main results. In Sec.~\ref{sec:semi}
we provide, for the interested readers, more details on our derivations and on the connection 
to the classical regime of Caron-Huot, while in 
Sec.~\ref{sec:LPM} we provide a pathway towards the investigation
of radiative corrections beyond the harmonic-oscillator and double-logarithmic
approximations. Finally, in Sec.~\ref{sec:concl} we draw our conclusions.
Our conventions, as well as more technical detail on the calculations, are 
collected in the appendices.

\section{The double-logarithmic phase space in the literature}
\label{sec:dlog}

As our starting point, let us call $\cc(\kp)$ the differential-in-transverse-momentum
scattering rate, also known as scattering kernel, i.e.
\begin{equation}
    \label{defcc}
    \cc(\kp) \equiv(2\pi)^2 \frac{d\Gamma}{d^2\kp},\,\qquad \qhat(\mu) = \int^\qhatmu  \frac{d^2\kp}{(2\pi)^2}\kp^2\,\cc(\kp)\,,
\end{equation}
where $\frac{d\Gamma}{d^2\kp}$ is the rate for the \textit{hard jet parton} of energy $\Ejet\gg T$ and momentum along $z$ to 
acquire $\kp$ transverse momentum --- see App.~\ref{sec_conv} for our conventions.

$\qhatmu$ is some process-dependent UV regularisation for the Coulomb logarithm in the leading-order 
scattering kernel, $\cc(\kp)\propto \kp^{-4}$ for $\kp\gg gT$. We shall treat $\qhatmu$ as a parameter,
without necessarily identifying it with the \emph{saturation scale} $Q_s^2=\qhat \lmed$ as 
done in \cite{Liou:2013qya}. In the case of QCD, asymptotic freedom would in principle make the integration
UV-finite, but it would include in $\qhat$
\emph{Moli\`ere scattering}~\cite{Moliere:1947zza,Moliere:1948zz}, i.e.  very large momentum transfers, which 
necessarily give rise to two hard jet partons in the final state, 
in what is no longer a diffusive process --- see e.g.~\cite{DEramo:2018eoy,Dai:2020rlu}.
This $\qhatmu$ factorisation also allows the incorporation in the effective kinetic description
of \cite{Arnold:2002zm,Ghiglieri:2015ala}, where $\qhat(\qhatmu)$ is (one of) the transport coefficients
describing diffusive momentum exchanges at scales below $\qhatmu$. There it is  complemented by 
the full kinetic description above that cutoff.

\begin{figure}[t]
    \begin{center}
        \includegraphics[width=0.32\textwidth]{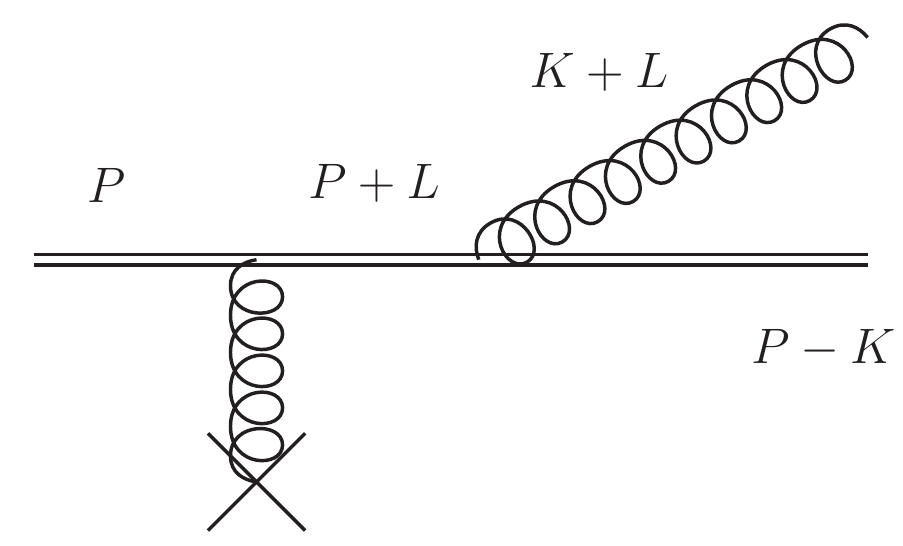}
        \includegraphics[width=0.32\textwidth]{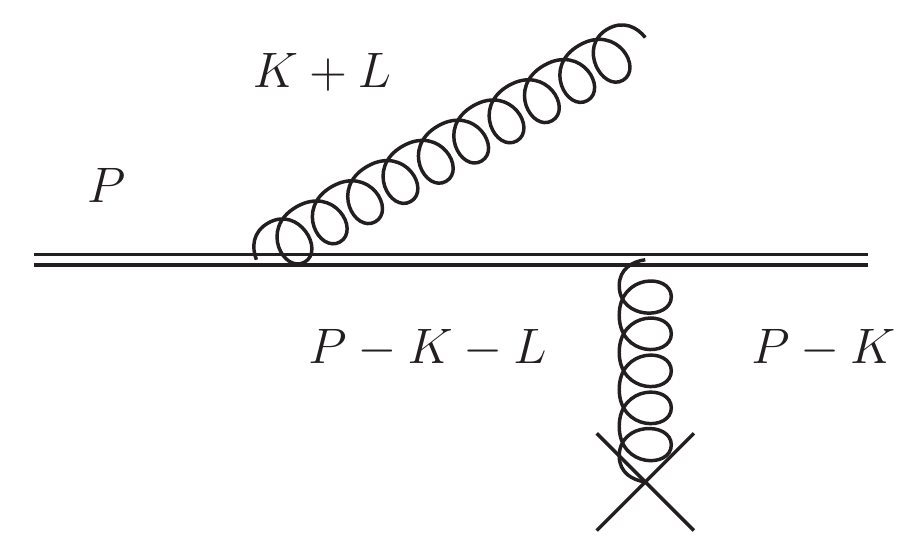}
        \includegraphics[width=0.32\textwidth]{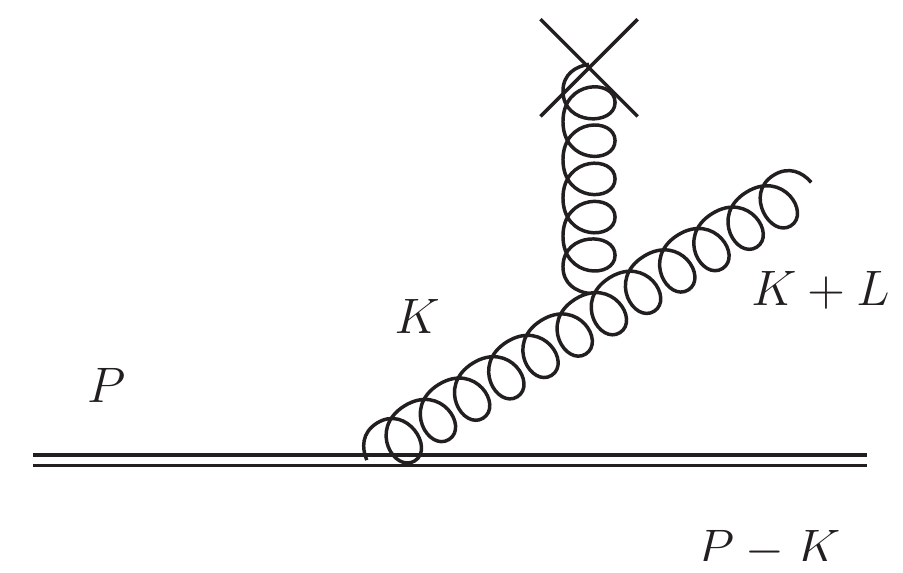}
    \end{center}
    \caption{Diagrams contributing to radiative momentum broadening 
    in the single-scattering regime. The double solid line represents 
    the hard jet parton (quark or gluon), curly lines are gluons and the 
    cross is a quark or gluon scattering center in the medium.}
    \label{fig:dl_diags}
\end{figure}
We now consider the radiative correction to the transverse 
scattering rate in the single-scattering regime,
which emerges naturally in the standard dipole picture \cite{Liou:2013qya}. It comes
from the diagrams in Fig.~\ref{fig:dl_diags} and it reads
\begin{equation}
    \label{N1term}
    \delta\cc(\kp)_\rad^{N=1}=4\als C_R\int\frac{dk^+}{k^+}\int \frac{d^2\lp}{(2\pi)^2}\cc_0(\lp)\frac{\lp^2}{\kp^2(\blp+\bkp)^2},
\end{equation}
where  $C_R$ is the Casimir 
factor of the  hard jet parton, $C_R=C_F=(N_c^2-1)/2N_c$ for a quark, $C_R=C_A=N_c$ for a gluon. 
As shown in Fig~\ref{fig:dl_diags}, $P$ is the four-momentum of the hard jet parton, with
$p^0=p^z=E\gg T$.
$L$ denotes the four-momentum acquired from the medium,
and $K$ is chosen in such a way
that the hard jet parton 
acquires a final transverse momentum of modulus $\kp$. $k^+\equiv(k^0+k^z)/2\approx k^0$ 
is the light-cone frequency --- see App.~\ref{sec_conv}
for our conventions. $\lp^2/(\kp^2(\blp+\bkp)^2)$ is then the standard dipole factor and
$2C_R/k^+$ the soft limit of the $g\leftarrow  R$ DGLAP splitting function. Finally, 
$\cc_0(\lp)\sim g^4 C_AT^3/\lp^4$ is the leading order scattering rate from a gluon source. In 
App.~\ref{app_scatt} we list the known LO results for $m_D\lesssim \lp \ll T$ and 
for $\lp\gtrsim T$, as well as a smooth interpolating scheme \cite{Aurenche:2002pd,Arnold:2008vd}.
For the present discussion the precise form is irrelevant, as we shall soon see.

Eq.~\eqref{N1term} is in principle just the first, $N=1$ term in the \emph{opacity series} of multiple scattering. 
As observed in \cite{Liou:2013qya}, the requirement that the transverse momentum carried away by the gluon is larger than that picked up 
in a typical collision with the medium, $\vert\bkp+\blp\vert\gg\lp$, puts us in the \emph{single-scattering regime},
where the $N=1$ term dominates. This yields 
\begin{equation}
    \label{ccHOapprox}
    \delta\cc(\kp)_\rad^\mathrm{single}=4\als C_R\,\qhatz\int\frac{dk^+}{k^+}\frac{1}{\kp^4}\,,\qquad\text{with}\quad 
    \qhatz \equiv \int^\homu \frac{d^2\lp}{(2\pi)^2}\,\lp^2\,\cc_0(\lp)\,,
\end{equation}
where we have also taken the \emph{harmonic-oscillator (HO) approximation}:  we  
neglect the $\homu$ scale dependence of the LO transverse momentum broadening coefficient
$\qhatz$, which is treated as a constant at the HO level. This furthermore makes 
the dependence on the detailed form of $\cc_0(\lp)$ irrelevant. We shall
discuss a pathway to go beyond this approximation in Sec.~\ref{sec:LPM}.

The double-logarithmic correction then follows in the form 
\begin{equation}
    \label{dlogexpand}
    \delta\qhat(\qhatmu)=4\als C_R\,\qhatz\int^\mu\frac{d^2\kp}{(2\pi)^2\kp^2}\int\frac{dk^+}{k^+}\,.
\end{equation}
Here we can quite explicitly see how the double log emerges --- one coming from the soft $dk^+/k^+$ divergence 
and the other from the collinear $d\kp^2/\kp^2$ divergence. 

\begin{figure}[t]
	\centering
	\includegraphics[width=0.7\textwidth]{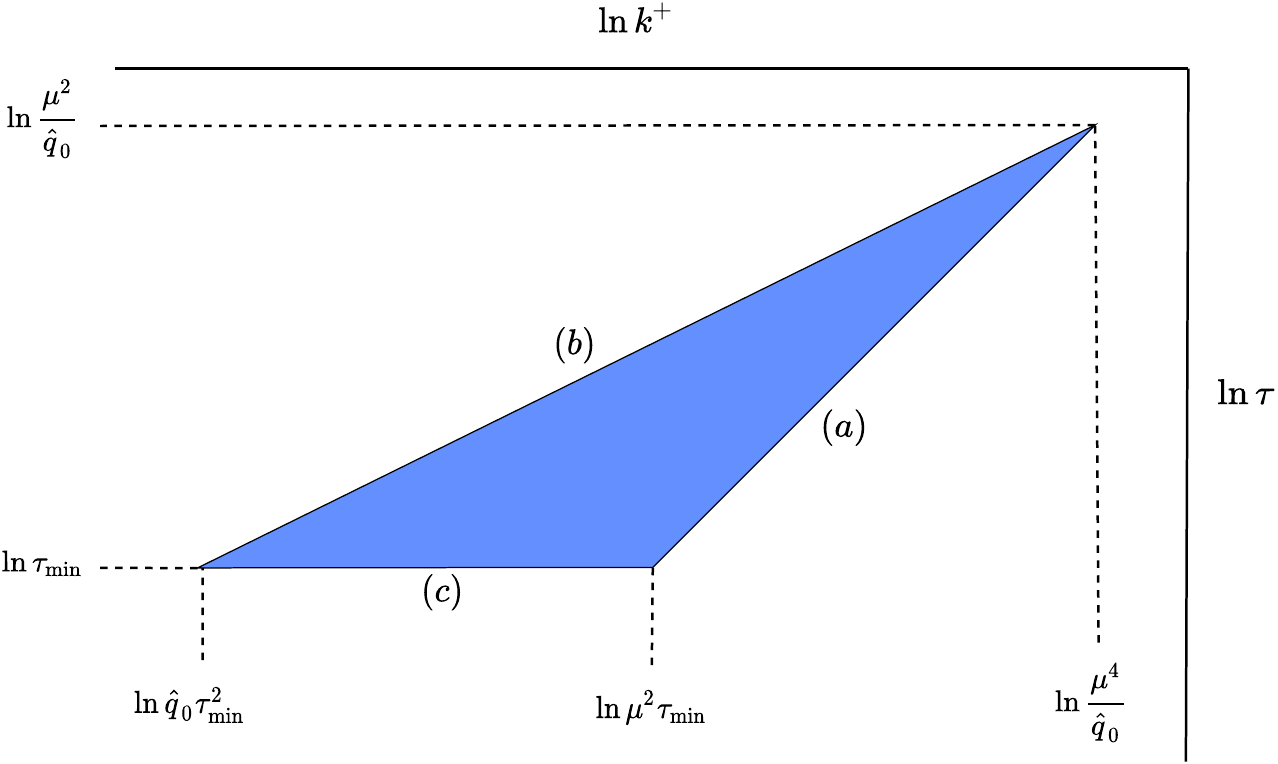}
	\caption{Schematic depiction of the bounds on the integral in Eq.~\eqref{eq:BDIMzeroT}. 
    The $(b)$ side of the triangle is given by $\ftime=\sqrt{k^+/\qhatz}$ whereas the
    $(a)$ one is given by $\ftime=k^+/\mu^2$. The logarithmic axes allow
    one to easily read the result of the integral Eq.~\eqref{eq:BDIMzeroT} straight from the figure, up to 
    the prefactor $\als C_R\qhatz/\pi$. Our labeling for the three boundaries follows that 
    of \cite{Liou:2013qya}.
    }
    \label{fig:bdimtriangle}
\end{figure}
We now specify the integration limits keeping us in the single scattering regime. Moreover, 
instead of integrating over $\kp$, it will be more convenient to integrate over the \emph{formation time} of the radiated gluon,
 $\ftime\equiv k^+/\kp^2$.  
The limits, which we show in Fig.~\ref{fig:bdimtriangle}, are then~\cite{Liou:2013qya}
\begin{itemize}
    \item $\ftime<\sqrt{k^+/\qhatz}$, represented by line $(b)$ in Fig.~\ref{fig:bdimtriangle}. 
    In the \emph{deep Landau--Pomeranchuk--Migdal (LPM)} multiple scattering regime, $\kp^2\sim\qhatz\ftime$. The former
    constraint then emerges upon demanding that $\ftime<\kp^2/\qhatz$ and solving for $\ftime$.
    Effectively, this prevents the formation time from getting sufficiently large to lead us into 
    the multiple scattering regime, which will cut off the double-logarithmic phase space:
    the collinear $dk_\perp^2/k_\perp^2$ log can exist as long as the initial and final
    states can propagate along straight lines for sufficiently long times 
    before and after the single scattering~\cite{Arnold:2008zu}.
   \item $\ftime>k^+/\qhatmu^2$, represented by line $(a)$. This condition on the formation time corresponds
    to enforcing the UV cutoff on transverse momentum, i.e. $\kp<\qhatmu$.   
    In the original derivation of \cite{Liou:2013qya} this $\qhatmu$ cutoff  is identified with 
    $Q_s^2\equiv \qhat \lmed$. 
    If instead $\qhatmu> Q_s$ the boundaries of the 
    double-logarithmic region change, as shown in \cite{Blaizot:2019muz}. 
    \item $\ftime>\tmin$, represented by line $(c)$. 
    This  IR cutoff $\tmin$ is intended to render 
    the end result finite. The motivation for this boundary comes from requiring that the 
    duration of the single scattering does not become
    comparable to the formation time of the radiated gluon, as the former is treated as instantaneous 
    within the collinear-radiation framework. In \cite{Liou:2013qya,Blaizot:2013vha} this duration was assumed  to be
    proportional to the inverse temperature, leading to $\tmin\sim 1/T$. We return to this 
    assumption in the next section. 
\end{itemize}

 This leads us to\footnote{This agrees with Eq.~(B.8) of \cite{Blaizot:2013vha}.
  The only difference between Eq.~\eqref{eq:BDIMzeroT} and their integral is that, while they cut off the energy of the radiated gluon
with the energy $\Ejet$ of the parent hard jet parton, we instead take $k^+<\qhatmu^4/\qhatz<\Ejet$, which is equivalent
to what is done in \cite{Liou:2013qya}. In any case, this will not have an impact
on the arguments that follow.} 
\begin{equation}
    \delta\qhat_{\text{\cite{Liou:2013qya,Blaizot:2013vha}}}(\qhatmu)=\frac{\als C_R}{\pi}\qhatz\int_{\tmin}^{\qhatmu^2/\qhatz}\frac{d\ftime}{\ftime}
    \int_{\qhatz\ftime^2}^{\qhatmu^2\ftime}\frac{dk^+}{k^+},
    \label{eq:BDIMzeroT}
\end{equation}
from which the double-logarithmic correction,
\begin{equation}
    \delta\qhat_{\text{\cite{Liou:2013qya,Blaizot:2013vha}}}(\qhatmu)=\frac{\als C_R}{2\pi}\qhatz\ln^2\frac{\qhatmu^2}{\qhatz\tmin}
    \label{eq:BDIMresultzeroT}
\end{equation}
immediately follows. This is the form of \cite{Blaizot:2013vha}; that of \cite{Liou:2013qya} arises by 
choosing $\qhatmu^2=\qhatz \lmed$, so that the argument of the double logarithm becomes the familiar $\lmed/\tmin$.
The single-logarithmic corrections arising from a more sophisticated analysis of the
regions neighboring the three boundaries of the triangle shown in Fig.~\ref{fig:bdimtriangle}
have been presented in \cite{Liou:2013qya}.

\section{Thermal scales in the double-logarithmic phase space}
\label{sec:temperature}

The double-logarithmic phase space we just sketched, as per \cite{Liou:2013qya,Blaizot:2013vha}, 
is derived  
for a medium 
described  by a random color field with a non-zero
two-point function for the $A^-$ component,\footnote{For a hard jet parton propagating in the $x^+$
light-cone direction.} i.e.
\begin{equation}
    \label{bdmimmedium}
    \left\langle A^{-a}(\bqp,x^+) A^{-b}(\bqp',x^{+'}) \right\rangle=\delta^{ab}\delta(x^+-x^{+'})
    n(x^+)(2\pi)^2\delta^{(2)}(\bqp-\bqp')\frac{g^4}{\qp^4}
\end{equation}
This corresponds for instance to the time-honored 
parametrisation of a medium of static scattering centres of density $n$. However, a weakly-coupled
QGP contains more medium effects than those captured by these instantaneous, space-like interactions; in 
particular, as soon as the light-cone frequency ($k^+$) range overlaps with the temperature scale, one needs
to account for the Bose--Einstein stimulated emission of the radiated gluon and, at negative $k^+$, for the 
absorption of a gluon from the medium. 
We could naively account for these effects by 
amending Eq.~\eqref{eq:BDIMzeroT} into 
\begin{equation}
    \delta \qhat(\qhatmu)=\frac{\als C_R}{\pi}\qhatz\int_{\tmin}^{\qhatmu^2/\qhatz}\frac{d\ftime}{\ftime}
    \int_{\qhatz\ftime^2}^{\qhatmu^2\ftime}\frac{dk^+}{k^+}\left(1+2\nbe(k^+)\right),
    \label{eq:BDIMnonzeroT}
\end{equation}
where 
    $\nbe(k^+)\equiv(e^\frac{k^+}{T}-1)^{-1}$
is the Bose-Einstein distribution. Its factor of two in Eq.~\eqref{eq:BDIMnonzeroT} accounts for 
stimulated emission and for absorption, which has been reflected in the positive-frequency range. 

The smallest frequency in Eq.~\eqref{eq:BDIMnonzeroT}
and in the corresponding original triangle of Fig.~\ref{fig:bdimtriangle} is $\qhatz \tmin^2$. 
To completely exclude the $k^+\sim T$ range and thus
make the thermal contribution exponentially small, one 
should then require $\tmin\gg \sqrt{T/\qhatz}\sim 1/(g^2T)$.
Would this be consistent with a single-scattering picture? To answer this,
let us recall a few relevant timescales in weakly-coupled QGPs.
Scattering processes exchanging $ \kp^2$ happen at a \emph{rate} --- 
see e.g. \cite{Arnold:2007pg,Arnold:2008zu} 
\begin{equation}
    \Gamma_{ \kp^2}\sim \frac{g^4T^3}{\max(\kp^2,m_D^2)},\quad\text{for}\; \kp\gtrsim g^2T.
\end{equation}
This arises from the  $\kp^{-4}$ form of $\cc(\kp)$. 
We have not shown the Coulomb logarithm of $ \kp^2$ over $m_D^2$,
with $m_D\sim gT$ the Debye (chromoelectric) screening mass --- 
see Eq.~\eqref{defmd}.\footnote{\label{foot_ho}%
Omitting this logarithm is consistent with the adopted 
harmonic-oscillator approximation.} This also means that a $\kp^2\gg (gT)^2$ amount may 
be accumulated either through a rarer hard scattering or through multiple softer ones. 

We further recall that the \emph{duration} of a scattering
process is $\OO( \kp^{-1})$. Hence, for the rarer scatterings with 
 $\kp\sim T$ the duration is indeed
$\OO(1/T)$, as per \cite{Liou:2013qya,Blaizot:2013vha}, 
whereas for the more frequent soft scatterings the duration is longer by $1/g$,
so that in principle more care is needed in drawing and approaching the 
$\ftime\gtrsim\tmin$ boundary of the phase space --- line $(c)$ in 
Fig.~\ref{fig:bdimtriangle} --- in a weakly-coupled QGP.

In addition to this latest consideration, 
$1/g^2T$ is the mean free time between the frequent soft,  $\kp\sim gT$
scatterings with the medium constituents. The dipole factor
in Eq.~\eqref{N1term} is complemented  in a thermal QGP
by thermal masses of order $gT$ and by the resummation of these 
frequent soft scatterings on this $1/g^2T$ timescale~\cite{Arnold:2001ba,Arnold:2002ja}. This suppresses the impact
of the non-perturbative ultrasoft scatterings and makes it so that 
$\ftime\gtrsim 1/g^2T$ for the multiple scattering regime,
thus answering negatively our previous question.
In more detail,
if we look at the LPM estimate $\ftime\sim \sqrt{k^+/\qhatz}$,
we have that $\ftime\sim (g^2T)^{-1}$ for $k^+\sim T$ and that 
it grows for growing $k^+$. In other words, for $k^+\sim T$ 
only scatterings with $\kp\sim gT$ have to be resummed 
in the multiple scattering regime, whereas 
for $k^+\gg T$ multiple soft scatterings and a single harder 
scattering can both occur within this parametrically longer formation 
time, giving rise in turn to a logarithmic enhancement 
that justifies the HO approximation. 
We recommend the pedagogical discussion in \cite{Arnold:2008zu}
for further clarifications. 

We also note that, for frequencies parametrically smaller
than the temperature, which $\tmin\sim1/T$ allows ($\qhat/T^2\sim g^4T$), we then have
$\sqrt{k^+/\qhatz}<1/g^2T$; this boundary $(b)$ too becomes 
ill-defined there. 
Furthermore, the factorisation into soft and collinear logarithms undergirding 
Eq.~\eqref{eq:BDIMzeroT} is predicated 
on a collinear expansion $k^+\gg k_\perp$, i.e. $\ftime k^+\gg 1$ \cite{Iancu:2014kga}.
This $\ln \ftime>-\ln k^+$ line in principle excludes parts of Fig.~\ref{fig:bdimtriangle}:
the $(b)-(c)$ vertex is below it, since $\qhat\tmin^3\sim g^4\ll 1$ for $\tmin\sim 1/T$.
The exclusion stretches to the $\ln \ftime=-\ln k^+$ line,
which crosses lines $(c)$ and $(b)$ if $\qhatmu\tmin>1$, i.e. $\qhatmu>T$.
If instead $\qhatmu<T$ it crosses lines $(a)$ and $(b)$. 

Where and how does the thermal distribution contribute to Eq.~\eqref{eq:BDIMnonzeroT}? 
How should we address the duration-dependent formation time boundary? 
How should we make sense of the deep LPM boundary $(b)$ if $k^+\ll T$ and of the collinearity 
boundary $\ftime k^+>1$? 
And finally, how does the transition 
to the classical corrections studied in \cite{CaronHuot:2008ni} take place? 
As the $\ftime k^+>1$ discussion suggests, the answer turns out to depend on the magnitude of $\qhatmu$
compared to the temperature. We start by discussing the case $\qhatmu<T$.
The discussion of this section is limited to the double-logarithmic correction 
in the HO approximation, though some of our more technical results, as presented 
in the subsequent sections, pave the way for a treatment beyond this accuracy.

\subsection{Double logarithms for $\qhatmu<T$}
\label{sub:smallmu}

\begin{figure}[t]
	\centering
	\includegraphics[width=0.8\textwidth]{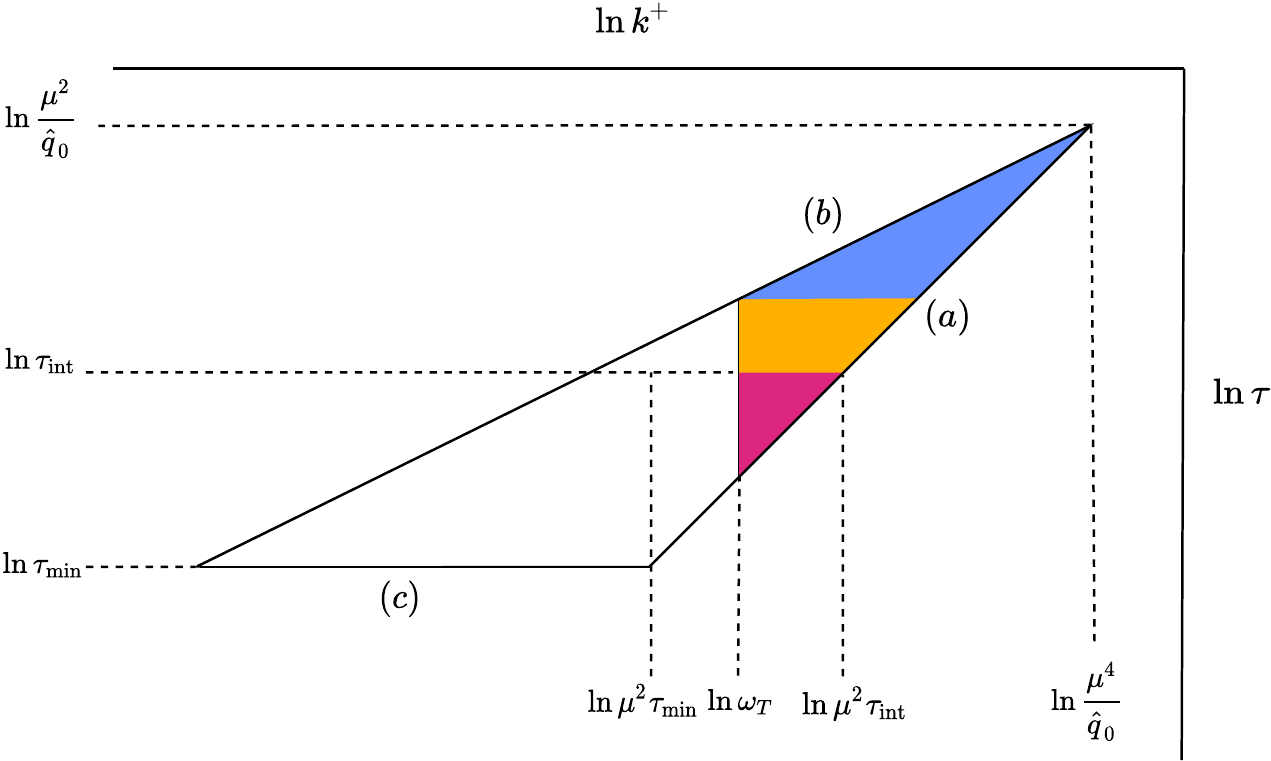}
    \put(-125,125){1}
    \put(-165,112){2}
    \put(-140,95){3}
    \put(-163,85){4}
    \put(-200,75){5}
	\caption{A pictorial representation of how the phase space of the radiated gluon is partitioned 
    once thermal effects are taken into account. See the main text 
    for the explanation of the different subregions.} 
    \label{fig:triangle}
\end{figure}
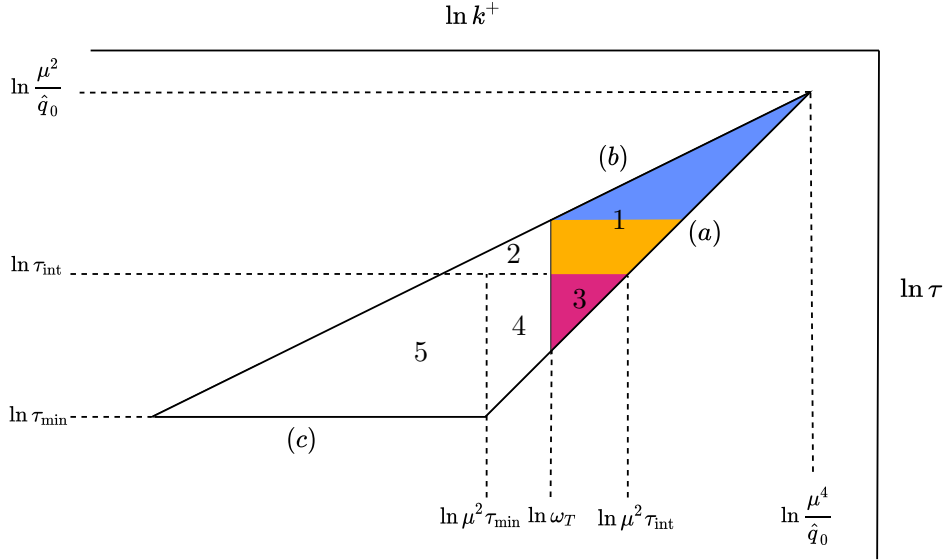

How can we address the issues we just outlined? A naive way out would be to require
$\tmin\gg 1/(g^2T)$, as depicted by the upper triangle in Fig.~\ref{fig:triangle},
which we have thus purposefully colored in the same shade as the original triangle
in Fig.~\ref{fig:bdimtriangle}: in this region $k^+ \gg T$ and the 
$\nbe$-proportional thermal contribution to Eq.~\eqref{eq:BDIMnonzeroT}
is exponentially small. 
However, as was previously noted, this large $\tmin$ would  amount
to only considering radiative processes with a long formation time, at odds 
with the $\ftime\ll 1/(g^2T)$ requirement for a \textit{strict} single soft scattering regime.\footnote{%
Multiple ultrasoft interactions, with $\lp\lesssim g^2T$, can occur
over this timescale, but these non-perturbative phenomena are suppressed 
by the dipole factor.}
Furthermore, as we shall explain, our detailed evaluation of the strict single scattering 
regime with $\ftime\ll 1/(g^2T)$, $k^+\sim T$ finds a double-log-enhanced contribution there.
We thus conclude that $\tmin$ cannot be larger than $1/g^2T$;
 for strict single scatterings to be included, we must have $\tmin\ll 1/(g^2T)$.
If line $(b)$ were to be taken seriously at $k^+<T$, we would 
thus be including parts of the thermal and subthermal frequency range. 

Moreover, our position is that the ``single-scattering'' definition employed in the literature on  
double logs
is not exclusively that of a strict single scattering: the collinear divergence gets cut off
when $\kp$ becomes small enough ($\ftime$ large enough) that multiple scatterings start to contribute
within a formation time, and that this happens before entering the so-called ``deep LPM'' 
regime $ \kp^2\sim \qhatz\ftime$.
However, at double-logarithmic accuracy we do not (in fact, we cannot) distinguish between a
$ \kp^2\gg \qhatz\ftime$ and a $ \kp^2\sim \qhatz\ftime$ boundary, and 
in the HO approximation --- see the discussion above and footnote~\ref{foot_ho} ---
we cannot disentangle multiple soft scatterings from a single harder one.
As mentioned, the analysis of \cite{Liou:2013qya} deals with single-logarithmic corrections 
to this boundary within the HO approximation;  in Sec.~\ref{sec:LPM}, 
we shall show how that boundary can be investigated in the future
beyond that approximation.

Our strategy is thus the following: 
we introduce an intermediate regulator $\tint$, 
with $1/gT \ll \tint\ll 1/g^2T$. The lower boundary will be discussed soon, whereas 
the upper boundary makes it so that for $\ftime >\tint$ 
we thus include parts of the strict single scattering regime,
as well as the ``few scatterings'' regime just mentioned. This region 
is then given by 
\begin{equation}
    \delta \qhat(\qhatmu)^{\mathrm{few}}=\frac{\als C_R}{\pi}\qhatz
    \int_{\tint}^{\qhatmu^2/\qhatz}\frac{d\ftime}{\ftime}
    \int_{\qhatz\ftime^2}^{\qhatmu^2\ftime}\frac{dk^+}{k^+}\left(1+2\nbe(k^+)\right).
    \label{eq:BDIMnonzeroTred}
\end{equation}

Graphically, it is represented in Fig.~\ref{fig:triangle} by the triangle above 
$\ftime=\tint$, which corresponds to the 1 and 2 subregions. We have filled the
1 region in blue for $\ftime\gtrsim 1/g^2T$, where the effect of the thermal distributions
is irrelevant, and in ochre for $\tint<\ftime< 1/g^2T$, where they start to contribute.
For reasons that will become clear shortly, the smaller 2 triangle has not been shaded.
Indeed, as we shall show in 
App.~\ref{app:integrals}, the double-logarithmic 
terms from the integration of Eq.~\eqref{eq:BDIMnonzeroTred} are
\begin{equation}
   \delta \qhat(\qhatmu)_\mathrm{dlog}^{\mathrm{few}}=\frac{\als C_R}{2\pi}\qhatz\bigg\{\ln^2\frac{\mu ^2}{\qhatz \tint} 
    -\frac{1}{2}\ln^2\frac{\omt}{\qhatz\tint^2}
    \bigg\}\quad \text{with}\;\omt=\frac{2\pi T}{e^{\gamma_E}}\quad \text{for}\;\;  
    \frac{\omt}{\mu^2}\ll\tint\ll\sqrt{\frac{\omt}{\qhatz}}.
    \label{eq:BDIMresultnonzeroT}
\end{equation}
Here $\gamma_E$ is the Euler--Mascheroni constant and $\omt$ is the $\OO(T)$ scale
 which naturally 
appears once thermal are effects into account. 
While its precise value can only be determined from the integration, its 
scaling could be expected and is related to the vacuum-thermal cancellation 
we shall discuss soon. Graphically, if we  take $k^+=\omt$ as a vertical line in Fig.~\ref{fig:triangle}, the first term
in Eq.~\eqref{eq:BDIMresultnonzeroT}
comes from integrating over the entire ``1+2'' triangle, whereas the second term comes
from subtracting the smaller 2 triangle, which therefore does not contribute to 
double-logarithmic accuracy. Furthermore, 
this $k^+=\omt$ line intersects line $(b)$ at $\sqrt{\omt/\qhatz}\sim 1/g^2T$, thus 
excluding the range where the formation time estimate becomes unreliable.

We also remark that, for $\tint>\tmin$, the horizontal $\ftime=\tint$ line
intersects the diagonal sides of the triangle at $k^+=\qhatz\tint^2$ (line $(b)$) and 
$k^+=\qhatmu^2 \tint$ (line $(a)$). The
form~\eqref{eq:BDIMresultnonzeroT} arises when the temperature scale $\omt$ falls in between these
two values, so that $\qhatz\tint^2\ll \omt \ll \qhatmu^2 \tint$, resulting in the range 
of validity expressed there. 
This is where the size of $\qhatmu$ with respect to the medium scales enters. 
At leading order, i.e. without considering these radiative corrections, choosing $gT\ll \qhatmu \ll T$
ensures that only the contributions of the soft modes \cite{Aurenche:2002pd} are included, whereas 
$T\ll \qhatmu$ also includes the thermal-mode contribution \cite{Arnold:2008vd}. If we allow 
for radiative corrections and further require a strict single-scattering contribution,
we shall show that $\qhatmu\gg \sqrt{g}T$ is necessary, so as to include 
the so-called \emph{semi-collinear processes} of \cite{Ghiglieri:2013gia,Ghiglieri:2015ala}.\footnote{%
\label{foot_small}
A good way of understanding why  $\kp\sim gT$ is not large enough,
thus moving us to the first available scaling $\kp\sim \sqrt{g}T$, $\qhatmu>\sqrt{g}T$, 
is that for $gT<\qhatmu<\sqrt{g}T$ the lines $(a)$ 
and $(b)$ intersect  at frequencies that are not parametrically 
larger than the temperature. The entire double-log triangle would 
then find itself at thermal or subthermal frequencies, effectively removing double-logarithmic enhancements.}
Finally, $\frac{\omt}{\mu^2}\ll\tint$ should be complemented by the duration bound $\tint\gg\tmin$: 
for $ \qhatmu\ll T$ this is automatically satisfied, whereas for $\qhatmu\gtrsim T$
it will then be a further constraint to be imposed. In other words,
for $\mu<\sqrt{\omt/\tmin}\sim T$ the vertical $k^+=\omt$ line intersects boundaries $(b)$ and $(a)$, 
whereas for $\mu>\sqrt{\omt/\tmin}\sim T$ it would intersect $(b)$ and $(c)$.

For formation times below $\tint$, the process happens in the strict single scattering regime. It 
also becomes sensitive to the fact that, over parts of that range, the \textit{duration} of
the strict single scattering, as per our previous discussion, 
is parametrically of the same order of the formation time. This happens because the 
region includes soft $\lp\sim gT$ scatterings whose duration overlaps 
with formation times of order $1/gT$, which are part of this region --- hence
our $\tint\gg 1/gT$ requirement.
As we shall show in Sec.~\ref{sec:semi}, our evaluation, based on these semi-collinear processes,
naturally accounts for these finite-duration
effects, removing the need for a hard cutoff $\ftime>\tmin$. 
This corresponds to the regions 3 and 4 in Fig.~\ref{fig:triangle}. 
As we mentioned, it also
 gives rise to a double-logarithmic correction.
It turns out that, to double-logarithmic accuracy in the HO approximation, 
our result corresponds to the area 
of the magenta 3 region --- we return to this fact soon.
It can thus be directly read off from Fig.~\ref{fig:triangle} as
\begin{equation}
    \delta\qhat(\qhatmu)^\text{single}_\mathrm{dlog}=\delta\qhat(\qhatmu)^\text{(3)}_\mathrm{dlog}=\frac{\als C_R}{2\pi}\qhatz\ln^2\frac{\qhatmu^2\tint}{\omt}.
\label{eq:redresult}
\end{equation}
Upon summing Eqs.~\eqref{eq:BDIMresultnonzeroT} and \eqref{eq:redresult} we then find
\begin{equation}
    \delta \qhat(\qhatmu)_\mathrm{dlog}=\delta \qhat(\qhatmu)_\mathrm{dlog}^{\mathrm{few}}+
    \delta\qhat(\qhatmu)^\text{single}_\mathrm{dlog}
    =\frac{\als C_R}{4\pi}\qhatz(\homu\ll\qhatmu)\ln^2\frac{\mu ^4}{\qhatz \omt}.
     \label{eq:dlogfinal}
 \end{equation}
We thus see that in this formula, which is one of our main results, 
$\tmin$ has disappeared: it has been replaced with the scale $\omt$. 
The dependence on the intermediate cutoff $\tint$ has also vanished.
We have further specified that the scale $\rho$ of $\qhatz$ needs to be smaller
than that of $\delta\qhat$, in keeping with the $\lp\ll \kp$ expansion
undergirding Eq.~\eqref{dlogexpand}. Since $\qhatmu\ll T$ here, the expression 
to be used is then Eq.~\eqref{qhatHOsoft}.
Eq.~\eqref{eq:dlogfinal} also justifies a posteriori --- see footnote~\ref{foot_small} --- that the first 
range of $\kp$ giving rise to a double-logarithmic enhancement is $\kp\sim \sqrt{g}T$, $\qhatmu>\sqrt{g}T$:
had we chosen $\kp\sim gT$ the argument of the double log in Eq.~\eqref{eq:dlogfinal}
would have been of order one.

Upon imposing $\qhatmu^2=\qhatz \lmed$, as in \cite{Liou:2013qya}, 
Eq.~\eqref{eq:dlogfinal} becomes
\begin{equation}
    \delta \qhat(\sqrt{\qhatz \lmed})_\mathrm{dlog}
    =\frac{\als C_R}{4\pi}\qhatz\ln^2\frac{\omc}{\omt},
     \label{eq:dlogfinalLMW}
 \end{equation}
where $\omc=\qhatz L^2$ is the maximal frequency for medium-induced radiation~\cite{Baier:1996kr}.
This formula is however only valid for $1/g^3T\ll\lmed\ll 1/g^4T$, that is for media thicker than a 
soft mean free path but thinner that a large-angle scattering mean free path.

We further remark that the effect of a populated medium can be understood, to 
double-logarithmic accuracy, as replacing the horizontal $\ftime>\tmin$ line 
with a vertical $k^+>\omt\sim T$ line: up to our usual prefactor of 
$\als C_R\qhatz/\pi$, $1/4\ln^2(\mu^4/\qhatz\omt)$ is precisely the area of the
``1+3'' shaded triangle in Fig.~\ref{fig:triangle}. Since lines 
$k^+=\omt$ and $(a)$ intersect at $\ftime = \omt/\qhatmu^2$, this
shaded triangle is entirely above the collinearity bound $\ftime k^+>1$.

The physical picture
behind the emergence of the $\omt$ scale is that 
for $k^+\gg T$ the thermal part $\nbe(k^+)$  of the phase space factor $1/2+\nbe(k^+)$
is exponentially small, whereas  
for $k^+<T$, the vacuum part --- $1/2$ --- 
cancels against the first quantum correction coming from the expanded $\nbe(k^+)$, i.e.
\begin{equation}
    \label{softbose}
    \nbe(k^+\ll T)=\frac{T}{k^+}-\frac12+\OO\left(\frac{k^+}{T}\right).
\end{equation}
Hence, the logarithmic part of the integral is unaffected in the UV, but it is no longer cut off in the IR by 
the boundary of the integration --- $\qhatz \tmin^2$ --- but rather by a quantity of order $T$, whose
precise value $\omt$ emerges as a property of the integral in Eq.~\eqref{eq:BDIMnonzeroT}.
This can be better seen from the following simpler single-logarithmic integral, with $\nuir\ll T\ll \nuuv$
\begin{align}
    \int^{\nuuv}_{\nuir}\frac{dk^{+}}{k^{+}}\Big(\underbrace{1}_\mathrm{vacuum}+\underbrace{2\nbe (k^{+})}_\mathrm{thermal}\Big)\nonumber
    &
    =\underbrace{\ln\frac{\nuuv}{\nuir}}_\mathrm{vacuum}\; 
    +\underbrace{\frac{2T}{\nuir}- \ln\frac{2\pi T}{\nuir e^{\gamma_E}}+\OO\left(\frac{\nuir}{T},\exp(-\nuuv/T)\right)}_\mathrm{thermal}\nonumber\\
    &=\frac{2T}{\nuir}+\ln\frac{\nuuv e^{\gamma_E}}{2 \pi  T} +\OO\left(\frac{\nuir}{T},\exp(-\nuuv/T)\right)\,. \label{eq:kplusint}
\end{align}
More details on the evaluation are provided in App.~\ref{app:integrals}.\footnote{%
Analogous cancellations between quantum vacuum and thermal corrections
in the soft regime, precisely related to the $\pm 1/2$ term in the 
expansion of the thermal distributions --- the plus sign applies to fermions --- 
are known in the literature: they are discussed briefly and in general terms in \cite{Heinz:1986kz}.
They  shift the \textit{Bethe logarithm} in the spectrum of heavy quarkonium 
from its $m\als^5\ln(m\als/(m\als^2))$ form in vacuum 
to a $m\als^5\ln(m\als/\omt)$ form in a thermal medium obeying
$m\als\gg T\gg m\als^2\gg m_D$, as shown in \cite{Brambilla:2010vq} ---
see also \cite{Escobedo:2010tu} for the analogous case of muonic hydrogen. 
$m\als$ is the typical transferred momentum or inverse Bohr radius in this Coulombic,
non-relativistic bound state and $m\als^2$ the typical binding energy. 
This is, in single-logarithmic form, precisely the same cancellation we observe:
as soon as the temperature becomes larger than the IR scale $m\als^2$ in the vacuum log,
 $m\als^2$ gets replaced by $\omt$.
Similar cancellations also appear in the \emph{power corrections}
to Hard Thermal Loops, which receive both vacuum and thermal contributions which
are separately IR-divergent but whose sum is IR-finite \cite{Manuel:2016wqs,Carignano:2017ovz}.}

The attentive reader will have noticed that this cancellation does 
not affect the \textit{classical} $T/k^+$ term,
 which is the largest in the IR, though it is not logarithmically divergent. Instead this term will yield a
contribution that is proportional to power laws of $\qhatmu$ and $\tmin$; in the example just shown,
this would be the $T/\nuir$ term. 
Power-law terms are in general not physical: they just represent a non-logarithmic
sensitivity to a neighboring region and must, for IR-safe quantities like $\qhat$, cancel with 
opposite power laws from said region. In our case, this neighboring
region is the one where both $L$ and $K$ are soft. In that region, the diagrams 
in Fig.~\ref{fig:dl_diags} are part of the calculation of Caron-Huot \cite{CaronHuot:2008ni}.
In terms of Fig.~\ref{fig:triangle}, this region has overlap with 
the ``4'' and ``5'' trapezoids.
As we shall show in detail in Sec.~\ref{sec:semi} and App.~\ref{app_arc},
these classical power law terms cancel against those coming from 
power-law corrections to the calculation of~\cite{CaronHuot:2008ni} that we shall derive. 
This undoubtedly shows in a non-trivial way how the 
classical and quantum regions are connected.

In summary, for $\sqrt{g}T<\qhatmu<T$ --- at double-logarithmic accuracy
we can use the $\ll$ and $<$ delimiters interchangeably --- 
our main result is that the shape of the double-logarithmic phase space changes
from that of Fig.~\ref{fig:bdimtriangle} to the shaded ``1+3'' one of Fig.~\ref{fig:triangle}: 
the horizontal line $(c)$, $\ftime>\tmin\sim1/T$, gets replaced by a vertical line
$k^+>\omt\sim T$. 
The fact
that the more sophisticated evaluation of the semi-collinear region, to be presented
in Sec.~\ref{sec:semi}, together with the regulated evaluation above the intermediate 
cutoff $\tint$, reproduces this simple form signals that effects analyzed in 
the semi-collinear evaluation, such as the necessary relaxation of the
instantaneous approximation, are not relevant at double-logarithmic 
accuracy in the HO approximation, as we shall show there.

\subsection{Extension to $\qhatmu>T$}
\label{sub:largemu}

Up until this point, we have taken $T\gg\qhatmu\gg\sqrt{g}T$, the first range where double-logarithm corrections appear. We now move on to consider what happens
if $\qhatmu\gg T$ in terms of the structure of the relevant portion of phase space. 
If we start from Fig.~\ref{fig:triangle} and proceed to increase $\qhatmu$, line $(a)$
gets shifted to the right, making the original $(a)-(b)-(c)$ triangle larger. As we 
argued previously, once $\qhatmu>\sqrt{\omt/\tmin}\sim T$ the vertical $k^+=\omt$ line intersects
lines $(b)$ and $(c)$ rather than $(b)$ and $(a)$. Hence, nothing would change 
for the part of the evaluation for $\ftime>\tint$, as given by Eq.~\eqref{eq:BDIMresultnonzeroT}.
On the other hand, the evaluation of the 
strict single scattering regime, for $\ftime<\tint$, needs to be amended to account for this. It 
must also account for the fact that for $\mu\gtrsim T$ we are thus including strict single
scatterings where $\lp$ is approaching or possibly exceeding the $T$ scale. In 
principle one would thus need to account for the fact that the coefficient 
of the leading-log (harmonic-oscillator) $\qhatz$ is not constant. As shown 
in \cite{Arnold:2008vd} and summarized in App.~\ref{app_scatt}, 
$\qhatz(\rho\ll T)$ and $\qhatz(\rho\gg T)$ differ at leading-log by
15\% for $N_f=3$ and $N_c=3$,  so in a first approximation we may neglect this effect. We leave 
a more precise evaluation of semi-collinear processes in this region to future work.

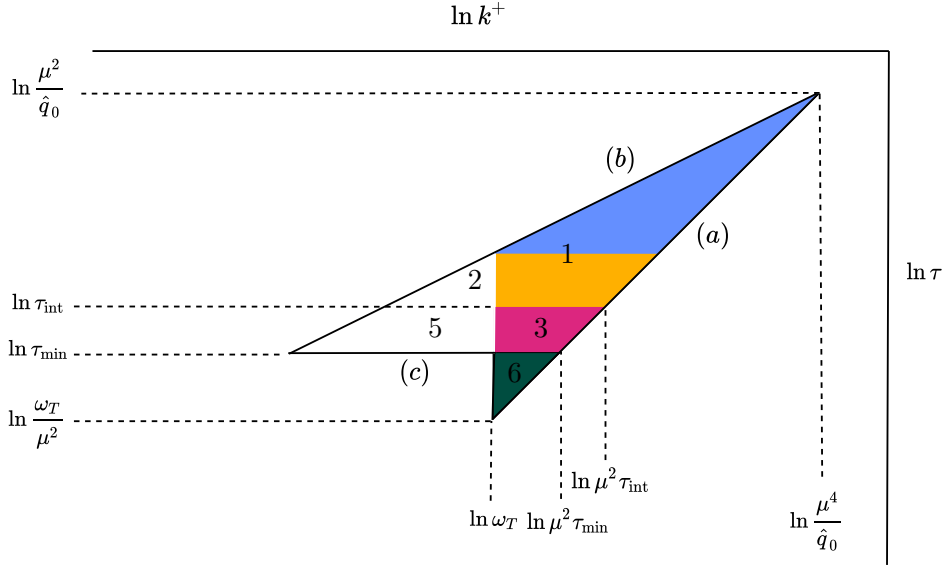
\begin{figure}[t]
	\centering
	\includegraphics[width=0.8\textwidth]{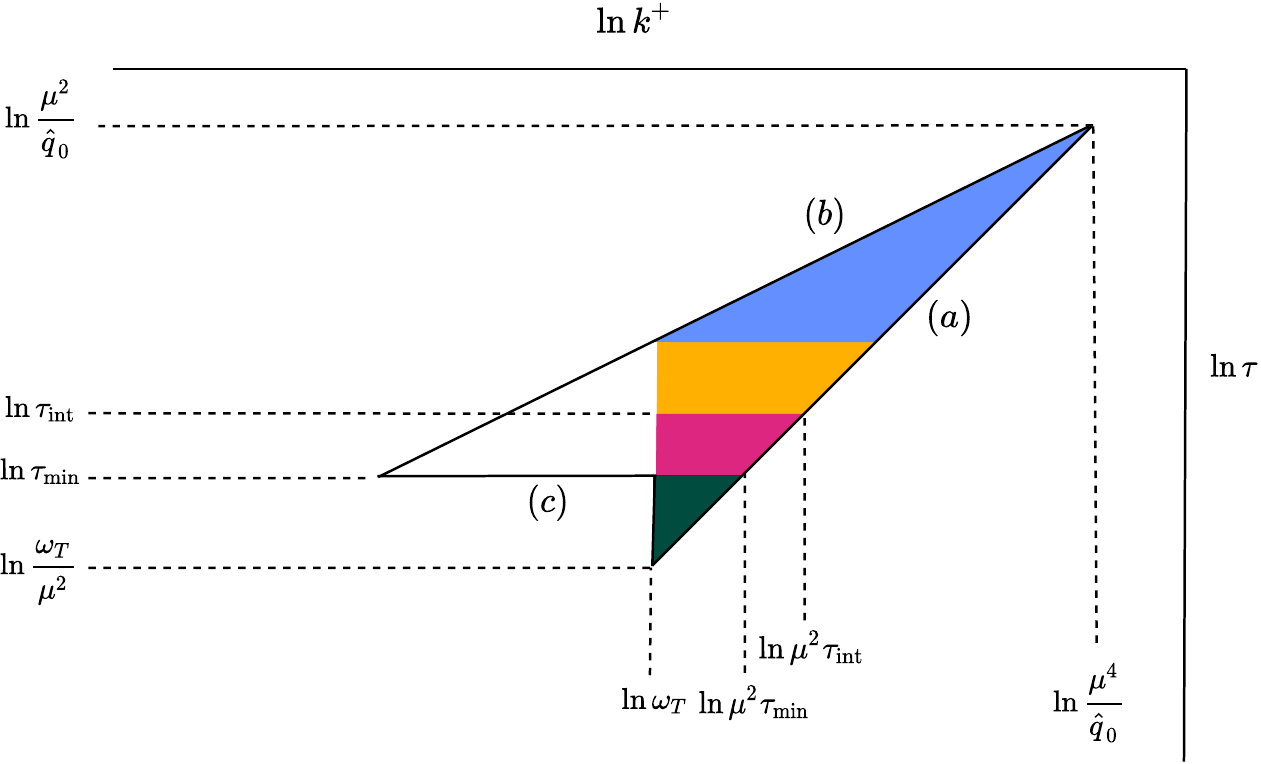}
    \put(-145,115){1}
    \put(-180,105){2}
    \put(-195,87){5}
    \put(-155,87){3}
    \put(-165,70){6}
	\caption{Structure of phase space once the cutoff on $\qhatmu$ is allowed to increase to 
    values much larger than the temperature. Notably, region ``6'' in dark green
    emerges. It cannot be well described by our setup due to the associated small formation time.}
    \label{fig:largemu}
\end{figure}
Under this approximation, we can simply extend Eq.~\eqref{eq:redresult} to the present case. 
This would however include the dark green triangle ``6'' in Fig.~\ref{fig:largemu},
which lies under the $\ftime=\tmin$ line. It is not clear whether this sharp line 
needs to be included, and only an evaluation like the one we show in detail in the next section,
but extended to larger $\qhatmu$ can address the effect of the relaxation of the instantaneous 
approximation in this setting. The expectation from the previous 
$\qhatmu<T$ results is that this effect should be subleading. We however choose
to proceed conservatively here and \emph{subtract} 
that slice from our result. Moreover, parts of this triangle lie below the $\ftime k^+=1$ line,
further motivating its subtraction.
In our double-logarithmic approximation this corresponds again,
for the reasons just explained in the previous subsection, to subtracting off the area of that triangle.
This leads to 
\begin{align}
    \delta \qhat(\qhatmu\gg T)_\mathrm{dlog}
    &=\frac{\als C_R}{2\pi}\qhatz(\homu\ll\qhatmu)\left[\frac12\ln^2\left(\frac{\mu ^4}{\qhatz \omt}\right)-
    \ln^2\left(\frac{\qhatmu^2\tmin}{\omt}\right) \right]\nonumber \\
    &=\frac{\als C_R}{2\pi}\qhatz(\homu\ll\qhatmu)\left[\ln^2\left(\frac{\mu ^2}{\qhatz \tmin}\right)-
    \frac12 \ln^2\left(\frac{\omt}{\qhatz \tmin^2}\right) \right] .
     \label{eq:dlogfinalhard}
 \end{align}
Unsurprisingly, the subtraction of the ``6'' triangle from the ``1+3+6''   triangle 
on the first line is equal to that of the unshaded ``2+5'' triangle from the original $(a)-(b)-(c)$ triangle
on the second line. This implies that, for $\qhatmu>T$, our result is closer to the original
one of \cite{Liou:2013qya}. This can be better appreciated by noting that the vertical line 
at $k^+=\qhatmu^2\tmin$ cuts the $(a)-(b)-(c)$ triangle into two triangles of equal area. Hence,
for $\omt>\qhatmu^2\tmin$, i.e. $\qhatmu\lesssim T$ our result~\eqref{eq:dlogfinal} is less than half of the original 
double logarithm~\eqref{eq:BDIMresultzeroT}, 
whereas for $\qhatmu>\sqrt{\omt/\tmin}$ our result~\eqref{eq:dlogfinalhard} is more than half of it.

\section{Single scattering for $\ftime<1/g^2T$, $k^+\gtrsim T$}
\label{sec:semi}
Our previous section presented all our main results to double-logarithmic accuracy
forgoing detail for the sake of a concise and self-contained explanation. In 
particular, we left out the detailed evaluation of our main computational 
result of this paper: the determination of the strict single 
scattering contribution for $\ftime<\tint$ and the connection 
to the soft, classical contribution. We now provide both. In Sec.~\ref{sub:setup}
we describe the general computational setup, in \ref{sub:semi} we introduce
semi-collinear processes and derive the double-log contribution.  In Sec.~\ref{sub:classical}
we discuss the interplay with the classical region and in Sec.~\ref{sub:subleading}
we analyze subleading contributions.
\subsection{Computational setup}
\label{sub:setup}
The natural computational setup  for a calculation of transverse momentum broadening
---  not  dissimilar from that of \cite{Liou:2013qya} --- 
is to
consider a Wilson loop in the $(x^+,x_\perp)$ plane --- see App.~\ref{sec_conv} for our conventions ---
through which we can define the scattering rate and $\qhat$ 
\cite{CasalderreySolana:2007qw,CaronHuot:2008ni,DEramo:2010ak,Benzke:2012sz}. 
The Wilson loop reads, taking a quark source for illustration
\begin{equation}
    \langle W(x_{\perp})\rangle=\frac{1}{N_{c}}\Tr\langle [0,x_{\perp}]_{-}\mathcal{W}^{\dagger}(x_{\perp})[x_{\perp},0]_{+}\mathcal{W}(0)\rangle, 
    \label{eq:Wloopdef}
    \end{equation}
where the Wilson lines read
\begin{equation}
    \mathcal{W}(x_{\perp})=\mathcal{P}\exp\Big(ig\int_{-\frac{\lmed}{2}}^{\frac{\lmed}{2}}dx^{+}A^{-}(x^{+},x_{\perp})\Big),
\end{equation}
and 
\begin{equation}
    [x_{\perp},y_{\perp}]_{\pm}=\mathcal{P}\exp\Big(-ig\int_{1}^{0}ds(y_{\perp}-x_{\perp})
    \cdot A_{\perp}(\pm \frac{\lmed}{2},x_{\perp}+(y_{\perp}-x_{\perp})s)\Big).\label{eq:side_rails}
\end{equation}
The fields in this Wilson loop are to be understood as transforming in the representation of 
the parton and  as path-ordered, so that one can think 
(in non-singular gauges) of $\mathcal{W}$ as the eikonalized jet quark in the amplitude
and $\mathcal{W}^\dagger$ as its conjugate-amplitude counterpart.

Exponentiation dictates that
\begin{equation}
    \lim_{\lmed \to \infty}\langle W(x_{\perp})\rangle=\exp(-\mathcal{C}(x_{\perp})\lmed), \label{eq:cdef}
\end{equation}
which we can use, together with
 \begin{equation}
    \mathcal{C}(x_{\perp})=\int \frac{d^{2}k_{\perp}}{(2\pi)^{2}}
    (1-e^{i\bxp\cdot\bkp})\mathcal{C}(k_{\perp}),\label{eq:cpos}
 \end{equation}
to recover $\mathcal{C}(k_{\perp})$. 
In a non-singular gauge --- we shall adopt later on in App.~\ref{app:diagrams} the strict Coulomb gauge --- the 
$[\ldots,\ldots]_\pm$ Wilson lines do not contribute to $\mathcal{C}(x_{\perp})$.
Eq.~\eqref{eq:cpos} further shows how 
$x_\perp$-independent diagrams --- those with no gluon exchange between the two
$\mathcal{W}$  --- will not contribute to $\mathcal{C}(x_{\perp})$  and thus to
$\qhat$; they only contribute to probability conservation by reducing the 
probability of acquiring no transverse momentum. 

\subsection{Semi-collinear processes}
\label{sub:semi}
\begin{figure}[t]
    \begin{center}
        \includegraphics[width=14cm]{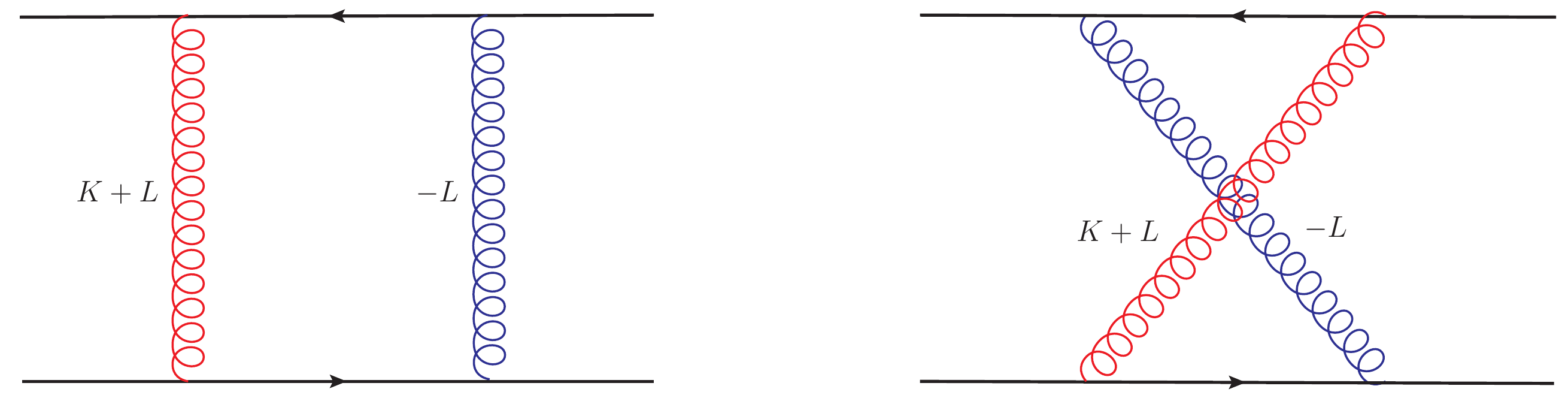}
        \includegraphics[width=5cm]{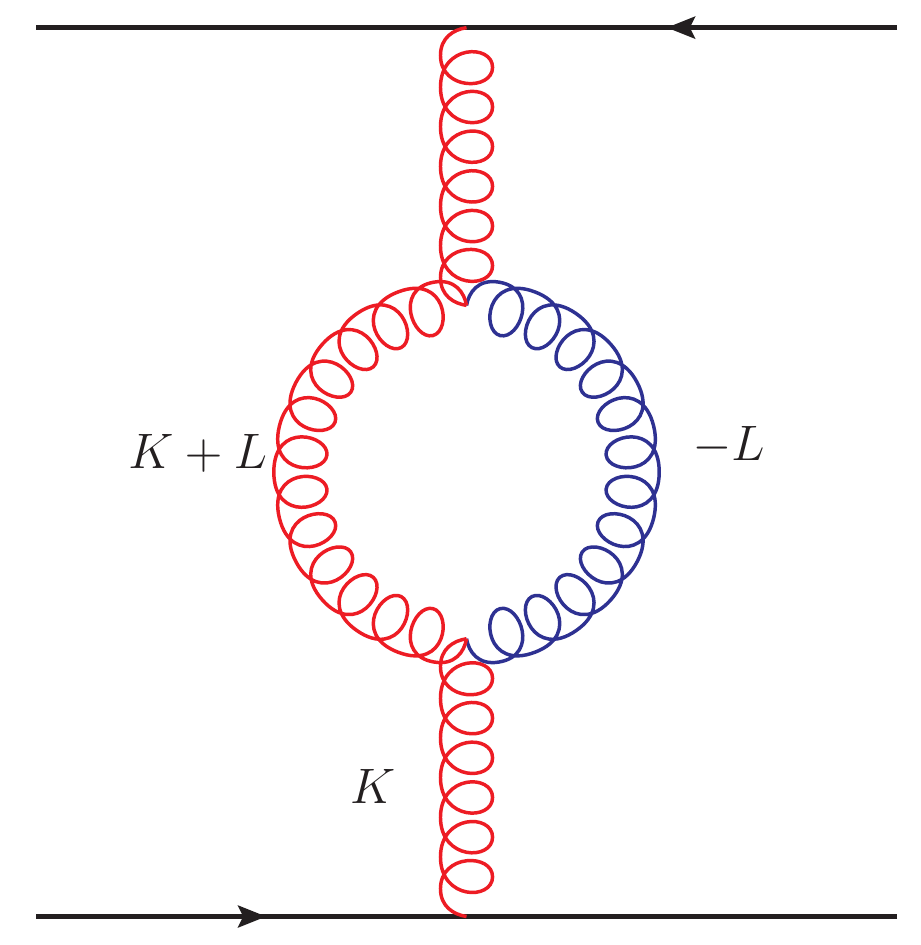}
        \includegraphics[width=7cm]{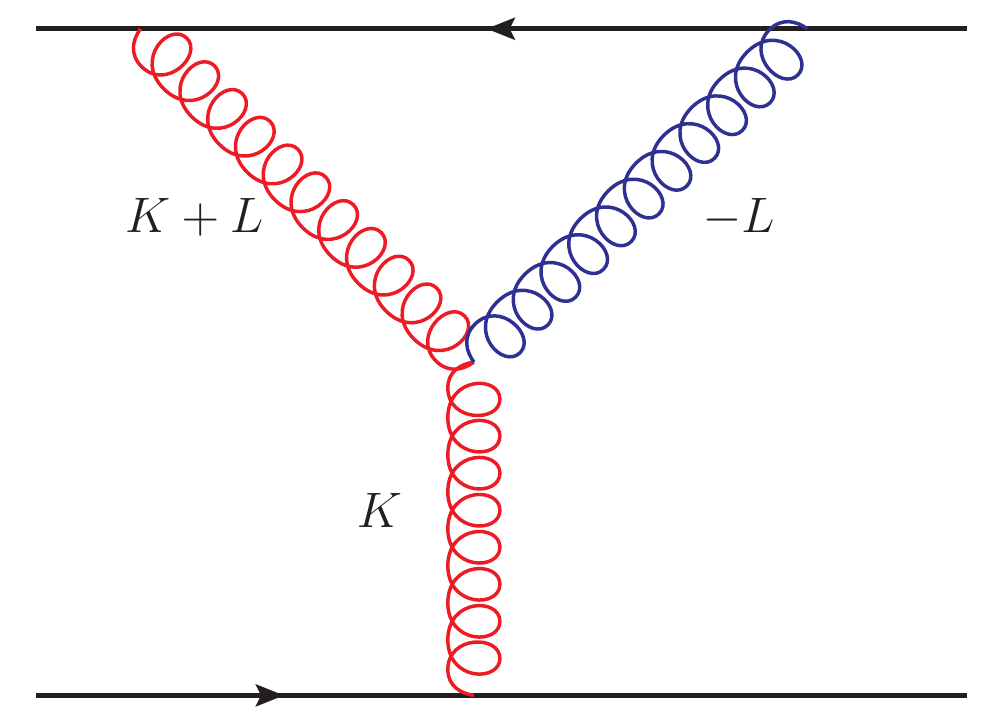}
    \end{center}
    \caption{Diagrams giving rise to the single-scattering
    regime. The two horizontal solid lines are the Wilson lines $\mathcal{W}$.
     Red gluons are \coft, blue ones are HTL-resummed and \soft, see main text.
     Momenta flow bottom to top.}
    \label{fig:real}
\end{figure}
Let  $P=(p^+,p^-,p_\perp)=E(1,0,0)$, with $E\gg T$, be the momentum 
of the hard jet parton. Then a mode $Q\sim E(1,\lambda^2,\lambda)$,
with $\lambda \ll 1$ an expansion parameter,
is a  collinear mode. 
\cite{DEramo:2010ak,Benzke:2012sz}
 argued that such modes are not included in this Wilson loop setup.
Radiative corrections like those we are after are on the
other hand included in this Wilson loop framework: the emitted gluon is collinear, but
it carries a momentum fraction $k^++l^+\approx k^+\ll E$. Borrowing the terminology 
of \cite{Becher:2015hka}, we call these modes \emph{\coft}, with $Q\sim \lambda^\prime E (1,\lambda^2,\lambda)$,
$\lambda,\lambda^\prime\ll 1$. The single-scattering processes
then arise, in the region we are interested in, from the interaction of these 
modes with HTL-resummed soft modes. We portray the relevant diagrams in Fig.~\ref{fig:real}.

As the two $\mathcal{W}$  lines represent the hard jet parton in the amplitude and 
conjugate amplitude, a cut is understood to go horizontally through the middle of each diagram. 
It then follows that the first two diagrams in Fig.~\ref{fig:real} corresponds to the square
of the first two in Fig.~\ref{fig:dl_diags} and their interference. The third diagram here
corresponds to the square of the third there. Finally the fourth here corresponds 
to the interference of the first two with the third there. 

So in principle we are presented with the evaluation of these diagrams in 
the specific scaling $\kp \sim \sqrt{g}T\gg \lp\sim gT$, $k^+\sim T$ that is responsible
for the double log in the strict scattering regime for $\mu<T$.\footnote{\label{foot_coft}%
This corresponds to the coft mode $K+L\sim T(1,g,\sqrt{g})$, i.e.
$\lambda^\prime=T/E,\lambda=\sqrt{g}$.} This detailed 
calculation will be presented in App.~\ref{app:diagrams}, together with the discussion
of the virtual counterpart
to the processes shown in Fig.~\ref{fig:dl_diags}, confirming 
that they do not represent a double-logarithmic effect, as per \cite{Liou:2013qya}. 
This Wilson-loop based determination will thus confirm how these radiative corrections
are also encoded in that object, thus paving the way for our later contact with the UV 
boundaries of the soft NLO calculation of Caron-Huot~\cite{CaronHuot:2008ni}, which 
also used the Wilson loop setup.

Here we instead present a more intuitive derivation, drawing from the literature. 
Namely,  $\kp \sim \sqrt{gT}\gg \lp\sim gT$, $k^+\sim T$ is the semi-collinear region
identified in \cite{Ghiglieri:2013gia,Ghiglieri:2015ala}. 
There it was found that it is 
precisely this \emph{semi-collinear process} that 
happens on a shorter formation time $\ftime_\semi\sim1/(gT)$ than the strictly collinear one
$\ftime_\mathrm{coll}\gtrsim1/(g^2T)$ through a single scattering with the medium.  The one single
scattering  exchanges $\lp\sim gT$; its duration
is thus of the same order of $\ftime_\semi$, causing the breakdown of the 
instantaneous approximation. Thus, addressing this 
corresponds to ``crossing''  boundary $(c)$, in the 
language of \cite{Liou:2013qya}.

\begin{figure}[t]
    \begin{center}
        \includegraphics[width=6cm]{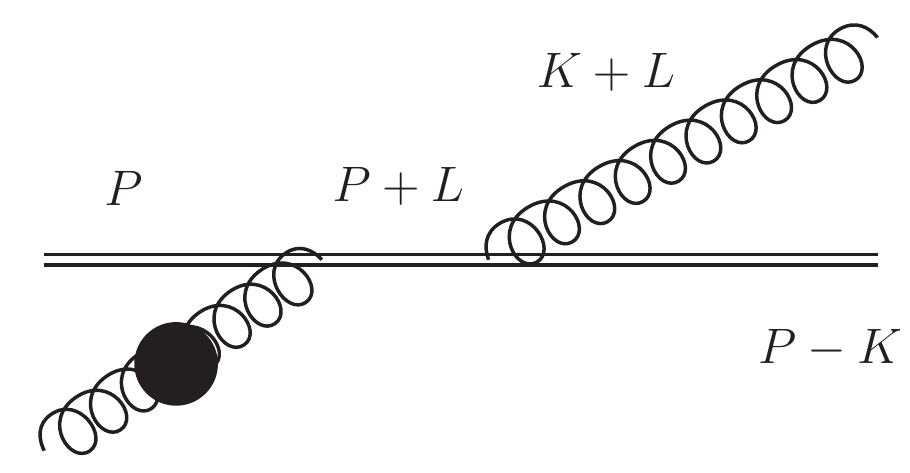}
    \end{center}
    \caption{One of the extra diagrams that appear once the duration
    of the jet-medium interaction is comparable to the gluon formation time. 
    The blob represents a resummed Hard Thermal Loop: the $L$ gluon is thus a time-like 
    plasmon. Diagrams corresponding to the other processes of Fig.~\ref{fig:dl_diags},
    as well as the plasmon emission/gluon absorption crossings are not shown.}
    \label{fig:semidiag}
\end{figure}

The evaluation
of \cite{Ghiglieri:2013gia,Ghiglieri:2015ala} addressed this non-instantaneous nature.
In more detail, the strict collinear regime corresponds,
in the momentum labeling of Fig.~\ref{fig:dl_diags}, to $l^-\approx(\bkp+\blp)^2/(2k^+)\sim g^2T\ll l^+,\lp\sim gT$,\footnote{%
We are taking the energy of the hard jet parton to be infinite,
in accordance with our general setup.} as 
arising from the on-shell conditions for the outgoing $K+L$ and $P-K$ legs. Thus, when 
Fourier-transforming the $L$ propagator to position space one can neglect its $l^-$ dependence,
leading to instantaneous propagation in the $x^+$ direction. Conversely, 
in the semi-collinear regime $\kp\sim\sqrt{g}T, k^+\sim T$, so that 
$l^-\approx \kp^2/(2k^+)\sim gT\sim l^+,\lp\sim gT$. Hence the outgoing 
gluon has the scaling $(K+L)\sim T(1,gT,\sqrt{g}T)$, which is what 
was identified in \cite{Ghiglieri:2013gia,Ghiglieri:2015ala} as 
semi-collinear. In our language it represents a specific \coft\ scaling, as per Footnote~\ref{foot_coft}.
In this scaling $l^-$ is no longer negligible with respect to $l^+$ and 
$\lp$, so that the $L$ gluon exchange is no longer instantaneous in 
$x^+$. In momentum space, this no longer restricts $L$ to space-like values, thus opening 
the phase space for the absorption and emission of soft, time-like plasmons, as shown  
in the example in Fig.~\ref{fig:semidiag}.

We can then directly take the results of \cite{Ghiglieri:2015ala}, which 
computed these semi-collinear processes in the non-abelian case. 
By inspecting Fig.~10 of \cite{Ghiglieri:2015ala} one arrives at the 
dictionary $K_{\text{\cite{Ghiglieri:2015ala}}}\to L$, $Q_{\text{\cite{Ghiglieri:2015ala}}}\to K$.
The derivation of \cite{Ghiglieri:2013gia,Ghiglieri:2015ala} lead to an 
integrated-in-transverse momentum rate; however, in intermediate steps a 
consistent labeling of momenta was maintained in all diagrams, so that 
this integration can be undone naturally. We can then take Eq.~(8.8) of 
\cite{Ghiglieri:2015ala}, undo the transverse integration, apply the dictionary and 
take the soft-gluon limit $x\to 0$ in the $q\to qg$ and $g\to gg$
processes, together with a $p\to\infty$ one for consistency.
This leads to
\begin{equation}
    \label{resultfromsemi}
    (2\pi)^2\frac{d\Gamma_\mathrm{semi}}{dk^+d^2\kp} = \frac{g^2 C_R  }{\pi k^+\kp^4}(1+\nbe(k^+))
    \qhat\left(\homu;\frac{\kp^2}{2k^+}\right),
\end{equation}
whence --- see Eq.~\eqref{defcc} 
\begin{equation}
    \label{resultfromsemi2}
    \delta\cc(\kp)_\mathrm{semi}= \frac{g^2 C_R}{\pi\kp^4}\int\frac{dk^+}{k^+}(1+\nbe(k^+))
    \qhat\left(\homu;\frac{\kp^2}{2k^+}\right).
\end{equation}
$\qhat(\homu;\kp^2/2k^+)$ is a modified (adjoint) $\qhat$ 
that also accounts for the $l^-$-dependence, i.e.
\begin{equation}
    \label{defqhatde}
   \qhat(\homu;l^-) =g^2 C_A T\int^\homu\frac{d^2\lp}{(2\pi)^2}
     \left[
         \frac{m_D^2\lp^2}{
         (\lp^2+l^{-2})(\lp^2+l^{-2} +m_D^2)}+2\frac{ l^{-2}}{\lp^2+l^{-2}} 
     \right].
\end{equation}
In the $l^-\ll gT$ limit it reduces to the soft contribution to $\qhatz(\homu)$ \cite{Aurenche:2002pd}
in Eq.~\eqref{qhatHOsoft}, and thus
Eq.~\eqref{resultfromsemi2} reduces, up to the statistical factor, to Eq.~\eqref{ccHOapprox}.
With respect to Eq.~(8.8) of \cite{Ghiglieri:2015ala} we have also undone the subtractions 
performed there, which were meant to isolate the strictly semi-collinear process from 
its collinear and harder limits, so as to avoid double countings. Something  
similar needs to happen here: 
elastic $2\leftrightarrow2$ scatterings exchanging  $\kp\gg gT$ are the hard
contribution to $\qhat$, as in \cite{Arnold:2008vd}. As we show in detail in 
App.~\ref{app:AX}, the integration over the leading-order phase space 
does include the region where $\sqrt{g}T\lesssim \kp\lesssim T$ and either
the incoming or outgoing gluon from the medium ($L$ here) becomes soft, $L\sim gT$.
As that calculation treats this gluon with bare propagators, it does not 
properly account for its soft dynamics, as encoded by HTL resummation. Hence, 
the semi-collinear limit of the calculation of \cite{Arnold:2008vd} must be 
subtracted from Eq.~\eqref{defqhatde}, to avoid double counting it. This yields
\begin{equation}
    \label{defqhatdesubtr}
    \qhat(\homu;l^-)_\subtr \equiv\qhat(\homu;l^-)-
    \qhat(\homu;l^-)_{\text{\cite{Arnold:2008vd}}} =g^2 C_A T
    \int^\homu\frac{d^2\lp}{(2\pi)^2}
         \frac{m_D^2\lp^2}{
         (\lp^2+l^{-2})(\lp^2+l^{-2} +m_D^2)},
\end{equation}
i.e. it removes the second term of Eq.~\eqref{defqhatde}, precisely as found in 
\cite{Ghiglieri:2013gia,Ghiglieri:2015ala}.

Eq.~\eqref{defqhatdesubtr}, when plugged in Eq.~\eqref{resultfromsemi2}, is UV log-divergent for $\rho\gg l^-,m_D$. 
This is not unexpected, as Eq.~\eqref{resultfromsemi2} is obtained under the assumption that $\lp\ll \kp$. We can thus 
set $\homu\ll \kp\ll\qhatmu$, leading to
\begin{equation}
    \label{qhatdecalc}
   \qhat(\homu;l^-)_\subtr
   = \als C_A T
   \bigg\{\underbrace{m_D^2 \ln\left(\frac{\homu^2}{m_D^2}\right)}_{\mathrm{HO}}\;
    \underbrace{-l^{-2}\ln\left(1+\frac{m_D^2}{l^{-2}}\right)-m_D^2\ln\left(1+\frac{l^{-2}}{m_D^2}\right)}
    _{l^--\mathrm{dependent}}
    \bigg\}
   .
\end{equation}
Our labeling in the underbraces emphasizes that the first, $l^-$-independent
term is precisely Eq.~\eqref{qhatHOsoft}, the 
harmonic-oscillator approximation to $\qhatz(\homu)$ for $gT\ll\rho\ll T$.
In our adoption of the HO approximation, we may for the moment treat $\homu$ as 
a parameter. If we were to go beyond it, we would have to complement 
the evaluation here with the neighboring region, where $\kp\sim\lp\gg gT$. 
That scaling includes both a single harder scattering and a multiple 
scattering contribution, which arises when $|\kp+\lp|$ becomes small,
causing the formation time to become long. Addressing these 
processes properly requires dealing with LPM resummation beyond the HO 
approximation: we discuss this outlook in Sec.~\ref{sec:LPM}.

Hence, the effect of our more sophisticated approach, which relaxes
the instantaneous approximation, is only contained in the $l^-$-dependent
terms in Eq.~\eqref{qhatdecalc}. The HO term is instead identical 
to what would have arisen directly from the simpler evaluation 
of Sec.~\ref{sec:temperature}. Its contribution to $\qhat$ is thus
\begin{equation}
    \label{semieval}
    \delta\qhat_\mathrm{semi}^\ho=
    4\als C_R \qhatz(\homu)
    \int\frac{d^2\kp}{(2\pi)^2}\frac{1}{\kp^2}\int\frac{dk^+}{k^+}(1+\nbe(k^+)),
\end{equation}
We can evaluate this expression by once again keeping only the even part of the frequency integrand 
and adding a factor of $2$ to account for the negative frequency range.
We can change variables to $\ftime=k^+/\kp^2=1/(2l^-)$ and 
enforce the boundaries  $\tint>\ftime>k^+/\qhatmu^2$, which are 
respectively the intermediate regulator and line $(a)$ of Sec.~\ref{sec:temperature}.
We further introduce an IR regulator for the frequency, $k^+>\kpir$, with $T\gg \kpir\gg gT$,
so as to avoid the soft frequency regime of \cite{CaronHuot:2008ni}.\footnote{
    \label{foot:kpir}
    We keep $\kpir$ as a generic cutoff satisfying $T\gg \kpir\gg gT$.
    The specific choice $\kpir=\mu^2\tmin$ would then precisely equate our 
    integration region with subregions 3 and 4 of Fig.~\ref{fig:triangle}.
}
This then implies 
$k^+<\qhatmu^2\tint$, i.e.
\begin{equation}
    \label{semievaltau}
    \delta\qhat_\mathrm{semi}^\ho= \frac{\als C_R }{\pi} \qhatz(\homu)
    \int_{\kpir}^{\qhatmu^2\tint}\frac{dk^+}{k^+}\left(1+2\nbe(k^+)\right)
    \int_{k^+/\qhatmu^2}^{\tint}\frac{d\ftime}{\ftime}.
\end{equation}
As we 
shall see, the $\kpir$ regulator only affects power law terms, but not double-logarithmic ones,
because of the vacuum-thermal cancellation discussed in Sec.~\ref{sec:temperature}. 
As we show in more detail in App.~\ref{app:integrals}, the integrations yield
\begin{equation}
    \label{semievaltaufinal}
    \delta\qhat_\mathrm{semi}^\ho= \frac{\als C_R }{2\pi} \qhatz(\homu)
    \bigg[ \frac{4T \ln \left(\frac{\qhatmu ^2
   \tint}{\kpir e}\right)}{\kpir}+
   \left( \ln
   ^2\left(\frac{\qhatmu ^2 \tint}{\omt}\right)-2 \gamma _1+\frac{\pi ^2}{4}- \gamma_E
   ^2\right)\bigg]+\OO\left(\kpir,e^{-\qhatmu\tint^2/T}\right),
\end{equation}
where $\gamma_n$ is the nth Stieltjes constant and $\omt=2\pi Te^{-\gamma_E}$. 
The double-logarithmic term is precisely the one anticipated 
in Eq.~\eqref{eq:redresult}. Here we  show the accompanying constant and the 
$T/\kpir$  power-law
divergence, which signals the overlap with the soft, classical region.

\subsection{Connection to the classical contribution}
\label{sub:classical}
\begin{figure}[t]
    \begin{center}
        \includegraphics[width=12cm]{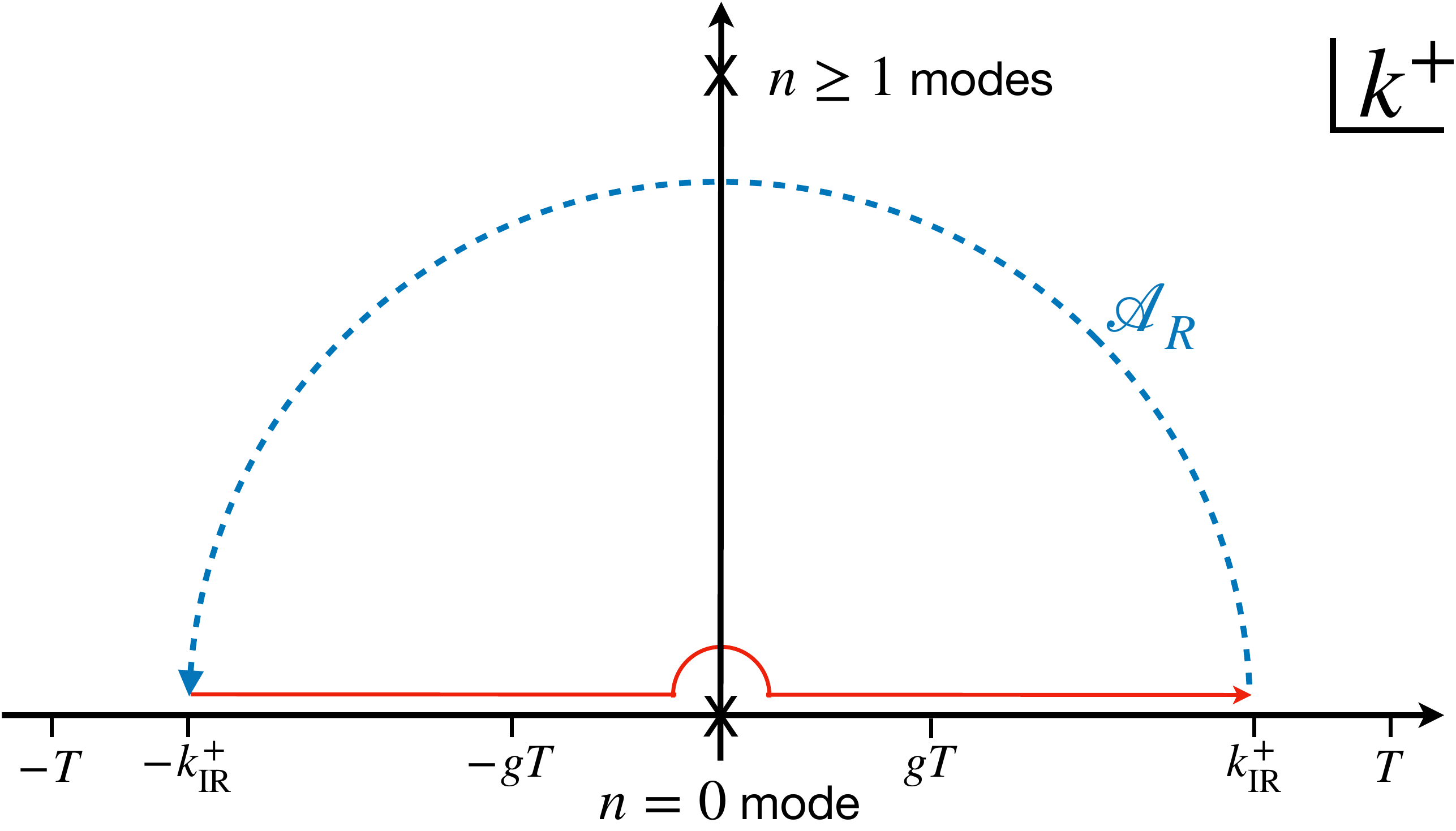}
    \end{center}
    \caption{The contour discussed in the main text 
    for the evaluation of the retarded part of the 
    soft contribution to Eq.~\eqref{treeAGZ}. The corresponding 
    advanced part is not shown. The cross at the origin is 
    Matsubara zero mode, hence the deformation of the (red) integration 
    contour there, which gives rise to the 
    Euclidean contribution.}
    \label{fig:matsubara}
\end{figure}
How does this $T/\kpir$ dependence cancel with the soft, classical region? Diagrams such as 
those in Fig.~\ref{fig:real} are precisely those evaluated 
in \cite{CaronHuot:2008ni} with $K\sim L\sim gT$. Hence, on that side
the calculation must present a $1/\kpir$ term, where now $\kpir$ is a 
UV cutoff that can thus be safely sent to infinity. Indeed, no 
such term appears in the expression of \cite{CaronHuot:2008ni}. Furthermore,
no $k^+$ integration appears either: as we mentioned,
the classical-mode contribution can be mapped to the 3D Euclidean theory --- see 
\cite{CaronHuot:2008ni} for the original derivation and 
\cite{Ghiglieri:2015zma} for a more pedagogical review. 
Let us look for illustration at the tree-level one-gluon exchange 
\begin{equation}
    \label{treeAGZ}
    \cc(\xp)=g^2 C_R\int\frac{d^4K}{(2\pi)^4}(1-e^{i\bkp\cdot\bxp})\left(
        \frac12+\nbe(k^0)\right)(G^{--}_R(K)-G^{--}_A(K))2\pi\delta(k^-),
\end{equation}
where the $\delta(k^-)$ comes from the Wilson-line integrations at large $\lmed$ 
and $G_R$ ($G_A$) is the retarded (advanced) gluon propagator.
The key observation is that causality dictates that $G_R$ ($G_A$) 
is analytical in the upper (lower) 
half $k^+$ plane, so that the only poles are those in the statistical 
function, at $k^0=k^+=i n2\pi T$, $n\in \mathbb{Z}$. In principle we can 
then close the contour at infinite $k^+$ and pick up the residues of 
all these poles. However, for $\xp\gg 1/T$ the zero mode dominates,
yielding the mapping to EQCD. This corresponds to having 
replaced $ 1/2+\nbe(k^+)$ with $T/k^+$ and closed the contour
on an arc $\mathcal{A}_R$ between the zeroth and first Matsubara modes, as shown in 
Fig.~\ref{fig:matsubara}. 
If we identify the radius of this 
arc with $\kpir$ (indeed $gT\ll \kpir\ll T$), we may take it as large, 
if we look at things from the soft side of the calculation. Hence, 
any function that falls to zero faster than $1/k^+$ on this arc 
will only give rise to inverse powers of $\kpir$, which could then safely
be neglected in the derivation of \cite{CaronHuot:2008ni}. 
This is indeed the case both for the LO and NLO soft contributions --- 
the LO case can be checked by plugging in Eq.~\eqref{treeAGZ}
the HTL-resummed Coulomb-gauge propagators given in Eqs.~\eqref{htllong} and \eqref{htltrans}.
In 
App.~\ref{app_arc} we show that \cite{Ghiglieri:2015ala}
computed diagrams 
related to those in Fig.~\ref{fig:real}, expanded precisely on that 
$\mathcal{A}_R$ arc. Starting from these results 
we show how the arc terms precisely cancel the $T/\kpir$ ones 
in Eq.~\eqref{semievaltaufinal}.

We have thus shown how the IR slice of the double-logarithmic phase space
overlaps with that of the soft contributions determined in \cite{CaronHuot:2008ni}.
In fact, \cite{CaronHuot:2008ni} already commented on the possible sensitivity 
to collinear modes (coft in our language) at relative order 
$g^2$ and how they would show up, on the soft side, 
as a failure of the $T/k^+$ classical approximation to the Bose--Einstein distribution.
Our findings thus confirm in detail this general expectation.

\subsection{Subleading contributions}
\label{sub:subleading}
We now turn to the remainder of Eq.~\eqref{qhatdecalc}. The $l^-$-dependent terms,
integrated with the same boundaries as in Eq.~\eqref{semieval},  give rise to  --- 
see again App.~\ref{app:integrals}
\begin{equation}
    \label{semievaltaufinalnon}
    \delta\qhat_\mathrm{semi}^{l^--\mathrm{dep}}= -\frac{\als^2 C_R C_A Tm_D^2 }{3\pi}\left[ 
    \ln^3\frac{e^{1/2}\mu^2}{2\omt m_D}+\mathcal{O}\left(\frac{T}{\kpir},\ln, \mathrm{const}\right)\right].
\end{equation}
We are not showing explicitly single-logarithmic terms and constants, which
are included in Eq.~\eqref{semievaltaufinalnontot}.  Neither do we show 
 the divergent power-law terms, proportional to $T/k^+$, which will 
again cancel against  classical contributions. We also do not show other 
power laws in the other cutoffs, as they can be similarly argued to cancel 
against neighboring regions.

We remark that the leading term in Eq.~\eqref{semievaltaufinalnon} is a \emph{triple logarithm} of $\mu^2/(m_D\omt)$. 
It is thus independent of $\tint$ and could then be directly added 
to Eq.~\eqref{eq:dlogfinal}. We argue that Eq.~\eqref{semievaltaufinalnon} 
is smaller than Eq.~\eqref{eq:dlogfinal}: the latter, reinstating the log 
of $\qhatz(\rho)$, is $\propto \ln(\homu^2/m_D^2)\ln^2(\qhatmu^2/\sqrt{\qhatz\omt})$. 
$\qhatmu^2/\sqrt{\qhatz\omt}$ is larger by $1/g$ than $\mu^2/(m_D\omt)$, which is instead
comparable to $\homu^2/m_D^2$, since $\homu\ll\qhatmu$. Hence Eq.~\eqref{semievaltaufinalnon}
represents a subleading correction; it is the first to feature a 
logarithmic dependence on the soft screening scale. This is somewhat 
analogous to what was found in \cite{Liou:2013qya} when crossing line $(c)$ 
--- and thus relaxing the instantaneous approximation --- for a nuclear medium:
it still generated a term with the highest number of logarithms, but with 
a smaller argument in some of them.

If we wanted to extend the present calculation to larger values of $\qhatmu$, 
so as to deal more precisely with region ``6'' of Fig.~\ref{fig:largemu},
we would need to understand how $\qhat(\homu;l^-)$ changes once $\homu$ 
starts to include the thermal range: this corresponds to the generalisation
to non-zero $l^-$ of the connection between the $\lp\sim gT$ and $\lp\sim T$ 
regimes discussed in App.~\ref{app_scatt}.
Let us finally point out that the subleading triple-log in Eq.~\eqref{semievaltaufinalnon}
is negative and that, for moderate values of the coupling $g$, 
may overtake the leading term in Eq.~\eqref{eq:dlogfinal}. And more generally,
the determinations at the highest logarithmic order (LL) are numerically not precise 
until the first (NLL) corrections are determined --- see for 
instance~\cite{Arnold:2008zu} for the splitting rate in the deep LPM 
regime and \cite{Arnold:2003zc} for transport coefficients.
Eq.~\eqref{semievaltaufinalnon} is only a part of the subleading corrections to 
the double-log in Eq.~\eqref{eq:dlogfinal}: in the next section 
we present a pathway to a more precise determination of boundary $(b)$.

\section{Transverse momentum broadening beyond the 
harmonic oscillator}
\label{sec:LPM}

Up until this point we have, with the exception of 
Sec.~\ref{sub:subleading}, only discussed double-logarithmic
corrections, and only done so within the harmonic-oscillator approximation.
Thus, we had to infer the coefficient of the HO $\qhatz$ from other 
considerations, i.e. the single-scattering requirement $\homu\ll\qhatmu$. 
We also explained in Sec.~\ref{sec:temperature} how, within these 
two approximations, we cannot distinguish the region where few scatterings 
are contributing from the deep LPM regime where many scatterings are contributing. 
Hence, the boundary $(b)$ had to be imposed by hand whenever relevant,
giving rise to the $\qhatz$ dependence in the argument of the double logarithm
in Eq.~\eqref{eq:dlogfinal}. In this section we now present, as an outlook,
a way to proceed beyond these two approximations and self-consistently
determine boundary $(b)$.

To this end, we will thus derive, starting from the formalism of~\cite{Liou:2013qya,Iancu:2014kga},
an LPM resummation equation that is not restricted to the harmonic-oscillator approximation.
By numerically solving that equation and performing the integrations over the two
logarithmic variables $k^+$ and $\ftime$ (or equivalently $k^+$ and $\kp$)
one would then see the emergence of multiple scatterings cutting off the 
double-log and thus be able to determine how good of an approximation line $(b)$
is.
Let us then start from Eqs.~(6), (7), (11) and (12) in \cite{Liou:2013qya} (see also (55)),
which construct a framework for resumming multiple interactions in the HO approximation
and in the large-$N_c$ limit. Combining (11) and (12) yields, in the notation of \cite{Liou:2013qya}
\begin{align}
    S(\xp)=-\als C_R\mathrm{Re}&\int \frac{d\omega}{\omega^3}\int_0^{\lmed}d z_2 \int_0^{z_2}dz_1 
    \nabla_{\bBpt}\cdot\nabla_{\bBpo}\nonumber\\
   &\hspace{-1cm} \times \bigg[e^{-\qhat_p \xp^2(\lmed-z_2+z_1)/4}
    G_\ho(\bBpt,z_2;\bBpo,z_1)
    -\mathrm{vac}
        \bigg]\bigg\vert^{\bBpt=\bxp}_{\bBpt=0}\bigg\vert^{\bBpo=\bxp}_{\bBpo=0},\label{LMWstart}
\end{align}
where we have already undone the large-$N_c$ approximation 
by replacing $N_c/2$, the original large-$N_c$ limit of $C_F$, with 
$C_R$. $\qhat_p$, with $p=q,g$ denotes the specific partonic broadening coefficients
for a quark or gluon source.
The $d\omega$ frequency integration is understood over the positive frequencies
of the radiated gluon.
The propagator $G_\ho$ is the 
Green's function of the following Schr\"odinger equation (see (55) there)
\begin{equation}
    \label{LMWschroHO}
    \bigg\{ i\partial_z +\frac{\nabla^2_{\Bp}}{2\omega}
    +\frac{i}{4}\bigg[\qhat_p \xp^2+\frac{\qhat_g}{2}\big(\Bp^2+(\bBp-\bxp)^2-\xp^2\big)\bigg]
    \bigg\}G_\ho(\bBp,z;\bBpo,z_1)=0,
\end{equation}
with 
\begin{equation}
    G_\ho(\bBp,z_1;\bBpo,z_1)=\delta^{(2)}(\bBp-\bBpo).
\end{equation}
``vac'' denotes the subtraction of the vacuum term $G_0$, i.e. the solution of Eq.~\eqref{LMWschro}
with a vanishing $\qhat$. The double vertical bars at the end 
of Eq.~\eqref{LMWstart} signify that the expression in brackets should be understood as
\begin{align}
    e^{-\frac{\qhat_p \xp^2}{4}(\lmed-z_2+z_1)}\bigg[ &G_\ho(\bxp,z_2;\bxp,z_1)+G_\ho(0,z_2;0,z_1)
   \nonumber\\
    &  -G_\ho(\bxp,z_2;0,z_1)-G_\ho(0,z_2;\bxp,z_1)\bigg]-\mathrm{vac}.  \label{LMWbars}
\end{align}
Physically, these four terms can be understood as the two positive, virtual terms, where the gluon
is emitted and reabsorbed within the amplitude (both $\bBp=0$) or conjugate amplitude (both $\bBp=\bxp$)
minus the real terms, where the gluon is emitted on one side and absorbed on the other side of the 
cut.

As a first step, we can 
identify their $S$-matrix element with
 our $\langle W(\xp)\rangle$.\footnote{To this end, it suffices
 to note that the broadening probability is given in our 
 framework by the Fourier transform of $\langle W(\xp)\rangle$ and  in 
 theirs by that of $S$ --- see their Eq.~(1).}
Hence we can equate 
\begin{equation}
S(\xp)=\exp[-\lmed(\mathcal{C}(\xp)+\delta\mathcal{C}(\xp))]\approx
e^{-\lmed\mathcal{C}(\xp)}(1-\lmed\delta\mathcal{C}(\xp)).
\label{eq:LMWdict}
\end{equation}
Secondly, Eq.~\eqref{LMWstart} is in the harmonic-oscillator approximation, i.e.
\begin{equation}
    \label{eq:HOexplicit}
    \mathcal{C}_{p}(\xp)\stackrel{\ho}{=}\frac{\qhat_p}{4}\xp^2,
\end{equation}
We can then undo this approximation, so that Eq.~\eqref{LMWschroHO} becomes
\begin{equation}
    \label{LMWschro}
    \bigg\{i\partial_z  +\frac{\nabla^2_{\Bp}-m_{\infty\,g}^2}{2\omega}
    +i\bigg[\cc_p(\xp)+\frac12\big(\cc_g(\Bp)+\cc_g(\vert\bBp-\bxp\vert)-\cc_g(\xp)\big)\bigg]
    \bigg\}G(\bBp,z;\bBpo,z_1)=0,
\end{equation}
where we have also introduced the gluon's asymptotic mass, with $m_{\infty\,g}^2=m_D^2/2$ at 
leading order --- see \cite{CaronHuot:2008uw} for the NLO determination and 
\cite{Moore:2020wvy,Ghiglieri:2021bom} for non-perturbative contributions.\footnote{\label{foot:mass}%
This mass term is necessary when $\nabla^2_{\Bp}$, the transverse momentum of the gluon,
becomes of order $g^2T^2$; it can be neglected in the deep LPM regime, where 
typical transverse momenta are larger, $\kp^2\sim\sqrt{\qhatz\omega}$, with $\omega\gg T$.}

Let us comment that the form of Eq.~\eqref{LMWschro} decomposes the three-body
scattering kernel (the hard jet parton in the amplitude and conjugate amplitude and the 
radiated gluon) into three two-body kernels with different color assignments. 
Perturbatively this is valid up to, and including, the $\OO(g)$ NLO corrections, as discussed 
in \cite{CaronHuot:2008ni}. The long-distance, non-perturbative
behaviour of this three-pole object is at present unknown. For a leading-order
determination of radiative correction from this formalism, it should suffice
to use the smooth kernel provided by the Fourier transform of Eq.~\eqref{hardqcdresum} as $\cc_q$ and 
$\cc_g$ in Eq.~\eqref{LMWschro}. We refer to \cite{Ghiglieri:2018ltw,Moore:2021jwe}
for details on this numerical transform.

Finally, we can also account for the effect of a populated medium by considering the effects of 
stimulated emission and absorption, i.e $\int d\omega\theta(\omega)\to \int dk^+ (1/2+\nbe(k^+))$ where we have 
replaced $\omega$ with $k^+$, in keeping with our notation. 
In a longitudinally uniform medium $G$ is only a function of $\ftime\equiv z_2-z_1$,\footnote{The 
generalisation to a longitudinally varying medium is straightforward.} so that we can
use our identification~\eqref{eq:LMWdict} to obtain, in the large-$\lmed$ limit
\begin{align}
    \delta \cc(\xp)=\als C_R\mathrm{Re}\int \frac{dk^+}{k^{+3}}\left(\frac12+\nbe(k^+)\right)
    \int_0^{\lmed}d \ftime \,
    \nabla_{\bBpt}\cdot\nabla_{\bBpo}\bigg[&e^{\cc_p(\xp)\ftime}
    G(\bBpt,\bBpo;\ftime)\nonumber\\
    &
   -\mathrm{vac}
        \bigg]\bigg\vert^{\bBpt=\bxp}_{\bBpt=0}\bigg\vert^{\bBpo=\bxp}_{\bBpo=0}.
        \label{LMWC}
\end{align}
We observe that the source-specific part of the scattering kernel in the Hamiltonian~\eqref{LMWschro} 
and the amplification factor $e^{\cc_p(\xp)\ftime}$ can be eliminated by noting that if
$\tilde G(\bBp,\bBpo;\ftime)\equiv e^{\cc_p(\xp)\ftime} G(\bBp,\bBpo;\ftime)$ is a Green's function of the operator
\begin{equation}
    \label{LMWschroshift}
    \bigg\{ i\partial_\ftime  +\frac{\nabla^2_{\Bp}-m_{\infty\,g}^2}{2k^+}
    +\frac{i}{2}\big(\cc_g(\Bp)+\cc_g(\vert\bBp-\bxp\vert)-\cc_g(\xp)\big)\bigg\}\tilde G(\bBp,\bBpo;\ftime)=0,
\end{equation}
then $G(\bBp,\bBpo;\ftime)$ is a Green's function of 
Eq.~\eqref{LMWschro}. Hence
\begin{align}
    \delta \cc(\xp)=\als C_R\mathrm{Re}\int \frac{dk^+}{k^{+3}}\left(\frac12+\nbe(k^+)\right)\int_0^{\lmed}d \ftime \,
    \nabla_{\bBpt}\cdot\nabla_{\bBpo}\bigg[&
    \tilde G(\bBpt,\bBpo;\ftime)\nonumber\\
    &
   -\mathrm{vac}
        \bigg]\bigg\vert^{\bBpt=\bxp}_{\bBpt=0}\bigg\vert^{\bBpo=\bxp}_{\bBpo=0}.
        \label{LMWCshift}
\end{align}
This reformulation makes transparent the fact that the Hamiltonian only contains 
the purely non-abelian $\cc_g$. That is because, before or after both
emission vertices, there are only the two source lines for the hard jet parton
 and one-gluon 
exchanges between them are resummed into $\exp\big(-\cc_p(\xp)(\lmed-\ftime)\big)$.
In the time region between the two emission vertices, the hard jet parton and 
conjugate hard jet parton lines are no longer a color singlet, with corresponding
color factor $C_R$, but rather an octet, with color factor 
$C_R-C_A/2$. This corresponds to the 
$\cc_p(\xp)-\cc_g(\xp)/2$ combination in Eq.~\eqref{LMWschro}. But we should 
remove the overall $\exp\big(-\cc_p(\xp)\lmed\big)$ damping, as per our 
dictionary~\eqref{eq:LMWdict},
which, thanks to our manipulation in Eq.~\eqref{LMWschroshift}, leads
to the outright disappearance of the $\cc_p(\xp)$ part, corresponding to the 
fact that those exchanges would happen also in the absence of the radiated 
gluon. Up to the statistical factors and thermal masses, Eqs.~\eqref{LMWschroshift} 
and ~\eqref{LMWCshift} agree with~\cite{Iancu:2014kga}.

Finally, for a medium that is isotropic in the azimuthal direction we can use 
the reflection symmetry $\bBp\to \bxp-\bBp$ of Eq.~\eqref{LMWschroshift}
to simplify Eq.~\eqref{LMWCshift} into
\begin{align}
    \delta \cc(\xp)=2\als C_R\mathrm{Re}\int \frac{dk^+}{k^{+3}}\left(\frac12+\nbe(k^+)\right)
    \int_0^{\lmed}d \ftime \,
    \nabla_{\bBpt}\cdot\nabla_{\bBpo}\bigg[&
    \tilde G(\bBpt,\bBpo;\ftime)\nonumber\\
    &
   -\mathrm{vac}
        \bigg]\bigg\vert^{\bBpt=\bxp,\bBpo=0}_{\bBpt=0,\bBpo=0}.
        \label{LMWCshiftfinal}
\end{align}

Eq.~\eqref{LMWschroshift} is, together with Eq.~\eqref{LMWCshiftfinal}, the main 
result of this section. As we anticipated, its solution would  allow 
a much better understanding of how the double logarithm is cut off by the transition from single
to multiple scatterings as a function of the energy $k^+$ 
of the radiated gluon. Furthermore, Eq.~\eqref{LMWschroshift} goes beyond the HO 
which we had to introduce in Eq.~\eqref{qhatdecalc}; as we remarked there,
regions where $\blp+\bkp$ becomes small would become sensitive to multiple scatterings, 
which are correctly addressed here.
However, Eq.~\eqref{LMWschroshift} is not easy to solve, as it would require 
generalizing the methods of \cite{CaronHuot:2010bp,Andres:2020vxs,Andres:2020kfg,Schlichting:2021idr}
to the extra dependence on $x_\perp$ of $\tilde{G}$. As a first step,
one could consider using the \emph{improved opacity expansion} introduced 
in~\cite{Mehtar-Tani:2019ygg,Barata:2020rdn,Barata:2021wuf,Isaksen:2022pkj} to capture 
the qualitative aspects of the transition from the HO approximation to 
including rarer harder scatterings.

We shall leave the full or approximate solution of Eq.~\eqref{LMWschroshift} to future work. 
We conclude this section by providing a non-trivial consistency check, namely that 
the single-scattering term in Eqs.~\eqref{LMWschroshift} and \eqref{LMWCshiftfinal} agrees with the standard 
results in the single-scattering regime. That follows by taking the 
Fourier transforms of Eqs.~\eqref{LMWschroshift} and 
Eq.~\eqref{LMWCshiftfinal}, i.e. starting from
\begin{equation}
    \tilde G(\bBpt,\bBpo;\ftime) \equiv \int\frac{d^2\pp}{(2\pi)^2}
    \int\frac{d^2\qp}{(2\pi)^2}e^{i\bBpt\cdot\bpp}e^{-i\bBpo\cdot\bqp}
    \tilde G(\bpp,\bqp;\ftime).
\end{equation}
Upon using the method of \cite{CaronHuot:2010bp} it is possible to 
obtain the $N=1$ contribution through a tedious calculation, yielding,
in the large-$\lmed$ limit
\begin{align}
    \delta \cc(\xp)^{N=1}=2\als C_R\int \frac{dk^+}{k^+}&\left(1+2\nbe(k^+)\right)\int\dpp\int\dlp \,
    \cc_g(\lp)\frac{(1-e^{i\bxp\cdot\bpp})(1+e^{i\bxp\cdot\blp})}
    {\pp^2+m_{\infty\,g}^2}\nonumber
    \\&\times\bigg[\frac{\pp^2}{\pp^2+m_{\infty\,g}^2}
    -\frac{\bpp\cdot(\bpp+\blp)}{
    (\bpp+\blp)^2+m_{\infty\,g}^2}
        \bigg].
    \label{eq:multiscat}
\end{align}

At this point we need to extract the real-process contribution, 
$\delta\mathcal{C}(k_{\perp})^\mathrm{real}_\mathrm{single}$.
This just corresponds to Fourier-transforming $\bxp$ into $\bkp$ and 
only keeping terms proportional to $\exp( i\bxp\cdot\bpp)$ and 
$\exp( i\bxp\cdot(\bpp+\blp))$, yielding
\begin{align}
    \delta\mathcal{C}(k_{\perp})^\mathrm{real}_\mathrm{single}
=&4\als C_{R}\int dk^+\frac{\frac12+\nbe(k^+)}{k^+}\int\frac{d^2l_\perp}{(2\pi)^{2}}
\cc_g(\lp) \left[\frac{\bkp}{\kp^2+m_{\infty\,g}^2}-
\frac{\bkp+\blp}{(\bkp{+}\blp)^2+m_{\infty\,g}^2}\right]^2.
\label{LWMsingle}
\end{align}
In the $m_{\infty\,g}\to0$ limit and neglecting the thermal distribution 
($1/2+\nbe(k^+)\to \theta(k^+)$)
this agrees with Eq.~\eqref{N1term}. The $\xp$-independent term 
in $\delta\cc(\xp)^{N=1}$ is the usual probability-conserving contribution,
whereas the term proportional to $\exp(i\bxp\cdot\blp)$ encodes
virtual processes. In App.~\ref{sub:virtual} we show its explicit 
form and prove that it agrees with our direct diagrammatic 
evaluation.

\section{Conclusions and outlook}
\label{sec:concl}
In this paper we have analyzed double-log enhanced quantum corrections 
to transverse momentum broadening in a weakly-coupled QGP. 
These corrections arise from the recoil in transverse momentum after 
a medium-induced radiation is sourced by a \emph{single scattering}
with the medium.
The region of
phase space for the double logarithm,  as represented 
in Fig.~\ref{fig:bdimtriangle}, was shown in~\cite{Liou:2013qya,Blaizot:2013vha} 
to be triangular in 
logarithmic units of formation time $\ftime$ and 
frequency $k^+$ of the radiated gluon, resulting
 in Eq.~\eqref{eq:BDIMresultzeroT}. We show that the boundary of said region 
 is only valid in media whose sole effect is 
to provide transverse-momentum kicks to the propagating hard jet parton. 

In our case, on the other hand, the thermal population of dynamical gluons changes 
the shape of the phase space, since the original triangle necessarily
overlaps with regions where $k^+$ is order $T$ or smaller. 
There one needs to account for Bose-stimulated emission 
and for absorption of thermal gluons from the bath.
This results in Figs.~\ref{fig:triangle} and \ref{fig:largemu},
which are the main results of this paper, together with the 
associated Eqs.~\eqref{eq:dlogfinal} and \eqref{eq:dlogfinalhard}.
What these figures and equations show is that thermal emission and absorption
change the infrared limits of the double-logarithmic integral:
no frequencies smaller than $\OO(T)$ contribute to these double logs.
This low-frequency region corresponds to the unshaded, white areas in both 
figures: only the colored regions to the right of the vertical $k^+=\omt=2\pi Te^{-\gamma_E}$
line contribute to the double logs. This is then reflected in the two
equations:  at double-logarithmic accuracy, the area of the shaded 
regions corresponds to the radiative correction.

How does this happen? We show that the effect of thermal absorption and emission
is described by changing the $dk^+/k^+$ logarithmic frequency integral into
$dk^+/k^+(1+2\nbe(k^+))$, with $\nbe$ the thermal distribution. As 
we explain in Sec.~\ref{sub:smallmu},
as soon as $k^+\ll T$, the IR log divergence in the vacuum part cancels
with twice the $-1/2$ from the IR expansion of the statistical function,
leaving the IR contribution to that integral to be dominated by the
non-logarithmic, classical $T/k^+$ term in that expansion. In 
Sec.~\ref{sub:classical} we show in detail how this term 
smoothly connects radiative quantum corrections to the 
classical soft corrections determined by Caron-Huot in \cite{CaronHuot:2008ni}
using the mapping to the three-dimensional Euclidean theory he introduced.
This is another of our main results: in a weakly-coupled quark-gluon plasma
the IR regions of the original double-logarithmic phase space are not $\OO(g^2\ln^2)$
quantum corrections but rather part of the $\OO(g)$ classical corrections. The
vacuum-thermal cancellation of the double-logarithmic piece in these regions
naturally separates the two corrections.

It is also worth stressing that our calculation identified two regions 
that contribute to double-logarithmic radiative corrections: taking 
Fig.~\ref{fig:triangle} as an example, for the case where the  
transverse momentum exchange is limited to $\qhatmu\lesssim T$, 
these are regions ``1'' and ``3''. We have identified  ``1''
as the region where single scatterings start to make way 
to a regime of ``few'' scatterings, before eventually reaching
the deep LPM, aka ``many scattering'' regime, represented by line $(b)$,
which is expected to cut off the double logarithm. Region ``3'' 
is instead restricted to formation times shorter than the mean free time
$1/g^2T$ between frequent soft scatterings in the medium: it is 
thus a strict single soft scattering regime, where furthermore
the duration of the single scattering overlaps with the formation time.
Our treatment in Sec.~\ref{sec:semi} is based on \emph{semi-collinear processes},
introduced in \cite{Ghiglieri:2013gia,Ghiglieri:2015ala}
and properly accounts for these overlapping timescales without 
resorting to instantaneous approximations. We however find
that the leading contribution from this region corresponds to 
what would emerge from a naive treatment which considers 
the single scattering instantaneous with respect to the formation time ---
see Sec.~\ref{sub:semi}. The effect arising from the overlapping
timescale, as given by Eq.~\eqref{semievaltaufinalnon}, though
still logarithmically enhanced, is subleading compared to 
the former, and represents part of the subleading log corrections
that are expected beyond double-logarithmic accuracy.

In Sec.~\ref{sec:LPM} we provide an important ingredient 
for the determination of radiative corrections beyond this accuracy:
we provide a framework --- Eqs.~\eqref{LMWschroshift} and  \eqref{LMWCshiftfinal} --- 
that can resum multiple scatterings 
and their LPM interference pattern beyond the harmonic-oscillator 
approximation, which, like previous literature, we have employed
for our previously discussed main results. This is obtained by
suitably extending the framework of \cite{Liou:2013qya,Iancu:2014kga}. Numerical 
or semi-analytical solutions of these equations will yield two 
important advancements: first, they would be able to 
shed light on how the transition from single to multiple scatterings
closes the double-logarithmic phase space at $k^+\gg T$, where 
in a weakly-coupled QGP the mean free time $1/g^2T$
between soft scatterings is well separated from the long formation time
$\ftime\sim \sqrt{k^+/\qhatz}\sim \sqrt{k^+/T}/g^2T$. Second,
if one wanted to approximate the solution with a double-logarithmic 
form, these solutions would clarify the $\homu$ scale at which $\qhatz$ should be evaluated 
in the harmonic-oscillator approximation and they would determine subleading
single-log corrections from boundary $(b)$. As these solutions
are not straightforwardly obtained, we leave these 
developments to future work.

This also implies that we are currently lacking a consistent determination
of all subleading single-log corrections.  We feel it would be premature and potentially
misleading to assess the quantitative
impact of these double-logarithmic corrections: in many cases --- 
see \cite{Arnold:2003zc} for transport coefficient and \cite{Arnold:2008zu}
for collinear splitting rates --- NLL corrections are necessary to have a sensible estimate
of the size of LL contributions. The situation is somewhat even more ambiguous in this 
case, as $\homu$ too is undetermined, as  explained.

Our results naturally open other directions for future research: as 
we discussed in the Introduction, these double- and single-logarithmic
radiative correction affect in the same fashion transverse momentum
broadening and double gluon emission \cite{Blaizot:2014bha,Wu:2014nca,Iancu:2014kga,Arnold:2021mow}. 
Addressing how the emergence of the temperature scale affects double gluon emission
would thus be a very  interesting natural development. 

Finally,  resummation equations for the double-log quantum corrections 
have been derived in~\cite{Liou:2013qya,Iancu:2014kga,Iancu:2014sha}
and solved in~\cite{Caucal:2021lgf,Caucal:2022fhc}. These methods 
are based on the original triangle of Fig.~\ref{fig:bdimtriangle}
for the double-logarithmic phase space and can be understood as 
evolving $\qhat$ from some initial timescale $\ftime_0$
to longer timescales by resumming many long-lived quantum fluctuations. 
A way to incorporate our careful evaluation of the thermal-$k^+$ and short-$\ftime$ regions
could be to use our results below some $\ftime_\mathrm{trans}\sim 1/g^2T$, properly
incorporating the vacuum-thermal cancellation as well as our semi-collinear single-scattering
regime and, potentially, also the classical regime with
its non-perturbative determinations \cite{Moore:2019lgw,Moore:2021jwe,Schlichting:2021idr}. 
The obtained $\qhat(\ftime_\mathrm{trans})$ could then be passed on 
as an initial condition to the resummation equations of~\cite{Caucal:2021lgf,Caucal:2022fhc},
thus naturally factorizing classical and semi-collinear contributions on 
one side from the quantum evolution on the other.\footnote{We are indebted to Paul Caucal for this proposal.}
This too is left to future investigations.

\section*{Acknowledgements}
JG and EW acknowledge support by a PULSAR grant from the R\'egion Pays de la Loire.
We are grateful to Paul Caucal for useful conversations.

\appendix

\section{Conventions}
\label{sec_conv}
Our sign for the covariant derivative is 
\begin{equation}
    D_{\mu}=\partial_{\mu}-igA_{\mu},\nonumber
\end{equation}
which fixes the sign of the three gluon vertex to be positive. 
Moreover, we work with the ``mostly minus" $(+,-,-,-)$ metric. 
Uppercase letters denote four-momenta, lowercase letters the modulus
of the three-momenta.

We will often be working in light-cone coordinates, where
\begin{align}
    p^{+}&\equiv\frac{p^{0}+p^{z}}{2}=\bar{v}_{\mu}p^{\mu},\nonumber
    \\p^{-}&\equiv p^{0}-p^{z}=v_{\mu}p^{\mu},\nonumber
    \\p\cdot q&=p^{+}q^{-}+p^{-}q^{+}-\bpp\cdot \bqp,\nonumber
\end{align}
where we have defined the two light-like reference vectors as 
\begin{align}
    \bar{v}^{\mu}&\equiv\frac{1}{2}(1,0,0,-1),\nonumber
    \\v^{\mu}&\equiv(1,0,0,1).\nonumber
\end{align}
This asymmetric convention for the ${}^+$ and ${}^-$ components of the light-cone 
coordinates has two advantages: it has unitary Jacobian, i.e. $dp^0dp^z=dp^+dp^-$,
and we shall often deal with scalings where $p^-\ll p^+$, which then implies
$p^0\approx p^z\approx p^+$.

Let us now discuss propagators. 
 For convenience
we will mostly work in the Keldysh, or $r,a$,
basis of the real-time formalism for the computation
of thermal expectation values --- see \cite{Ghiglieri:2020dpq} for a review. 
The two elements of this basis are defined as
$\phi_r\equiv(\phi_1+\phi_2)/2$,
$\phi_a\equiv\phi_1-\phi_2$, $\phi$ being a generic field and the subscripts 1 and 2
labeling the time-ordered and anti-time-ordered branches of the Schwinger-Keldysh contour respectively.
The propagator is a $2\times2$ matrix, where one entry is always zero and only one entry depends on the
thermal distribution, \emph{i.e.},
\begin{equation}
\label{raprop}
D=\left(\begin{array}{cc} D_{rr}&D_{ra}\\D_{ar}&D_{aa}\end{array}\right)
=\left(\begin{array}{ccc} \left(\frac12\pm n(p^0)\right)(D_R-D_A)&&D_{R}\\D_{A}&&0\end{array}\right),
\end{equation}
where $D_R$ and $D_A$ are the retarded and advanced propagators, the plus (minus) sign refers
to bosons (fermions). $n(p^0)$ is the corresponding thermal distribution, either
$\nbe(p^0)=(\exp(p^0/T)-1)^{-1}$ for bosons or $n_\mathrm{F}(p^0)=(\exp(p^0/T)+1)^{-1}$ for fermions.
We also define the spectral function as the difference of the retarded and
advanced propagators, $\rho \equiv D_R - D_A$.
We will denote the gluon propagator by $G$. 

We will adopt strict Coulomb gauge throughout.  The treatment of soft
momenta in propagators and vertices requires the use of Hard Thermal
Loop (HTL) resummation \cite{Braaten:1989mz}. 
Coulomb-gauge HTL-resummed gluons are described  by
\begin{eqnarray}
\label{htllong}
G^{00}_R(Q)&=&\frac{i}{\displaystyle q^2+m_D^2\left(1-\frac{q^0}{2q}\ln\frac{q^0+q+i\epsilon}{q^0-q+i\epsilon}\right)},\\
\nonumber G^{ij}_R(Q)&\equiv&(\delta^{ij}-\hat q^i\hat q^j)G^T_R(Q)=
\left.\frac{i(\delta^{ij}-\hat q^i\hat q^j)}
     {\displaystyle q_0^2-q^2-m_\infty^2 \left(\frac{q_0^2}{q^2}
       -\left(\frac{q_0^2}{q^2}-1\right)\frac{q^0}{2q}
       \ln\frac{q^0{+}q}{q^0{-}q}\right)}\right\vert_{q^0=q^0+i\epsilon}.\\
&& 	\label{htltrans}
\end{eqnarray}
Here 
\begin{equation}
    m_D^2=g^2 T^2\left(\frac{N_c}{3}+\frac{N_f}{6}\right)
    \label{defmd} 
\end{equation}
 is the LO Debye mass and $m_\infty^2=m_D^2/2$ is the LO
gluon asymptotic mass.
The other components of the propagators in the $r,a$ basis can be obtained
through Eq.~\eqref{raprop}.

\section{Transverse-momentum broadening kernels}
\label{app_scatt}
The leading-order transverse momentum broadening kernel comes
from elastic $2\leftrightarrow 2$ scatterings with medium constituents. In the soft
sector, i.e. $m_D\lesssim \lp\ll T$, these get Landau-damped, leading to 
\cite{Aurenche:2002pd}
\begin{equation}
    \label{losoft}
    \cc(\lp)^\mathrm{LO}_\mathrm{soft}=\frac{g^2 C_R T m_D^2}{\lp^2(\lp^2+m_D^2)}.
\end{equation}
For $\sqrt{ET}\gg\lp\gtrsim T$ one finds
 \cite{Arnold:2008vd}
\begin{equation}
\cc(\lp)_{\text{hard}}^\mathrm{LO}=
\frac{g^4C_R}{\lp^4}\int\frac{d^3q}{(2\pi)^3}\frac{q-q_z}{q}
\Big[2C_A\,\nbe(q)(1+\nbe(q'))
+4 N_fT_F\,\nfd(q)(1-\nfd(q'))\Big],
\label{hardqcd}
\end{equation}
where $q'=q+\frac{l_\perp^2+2\blp\cdot \bq}{2(q-q_z)}$ and $T_F=1/2$ for the quark scattering contribution. 
We refer to \cite{Arnold:2008vd} for details on the accurate
numerical evaluation of this expression. 
\cite{Arnold:2008vd} also provided a handy expression that interpolates smoothly
between the two regimes by partially resumming higher-order contributions, i.e.
\begin{equation}
    \cc(\lp)_{\text{smooth}}^\mathrm{LO}=
    \frac{2g^4C_R}{\lp^2(\lp^2+m_D^2)}\int\frac{d^3q}{(2\pi)^3}\frac{q-q_z}{q}
    \Big[C_A\,\nbe(q)(1+\nbe(q'))
    +2 N_fT_F\,\nfd(q)(1-\nfd(q'))\Big].
    \label{hardqcdresum}
\end{equation}
It is straightforward to show that this form reduces to Eq.~\eqref{losoft}
for $\lp\ll T$, whereas for $\lp\gg T$ it goes into
\begin{equation}
    \cc(\lp\gg T)_{\text{smooth}}^\mathrm{LO}=
    \frac{2g^4C_R}{\lp^4}\int\frac{d^3q}{(2\pi)^3}
    \Big[C_A\,\nbe(q)
    +2 N_fT_F\,\nfd(q)\Big]= \frac{g^4C_R\zeta(3)T^3}{\pi^2\lp^4}
    \Big[2C_A +3 N_f T_F\Big].
    \label{hardqcdresumhard}
\end{equation}
In between these two limits $\cc(\lp)^\mathrm{LO}_\mathrm{smooth}$ is a monotonic function.

The leading-log coefficients of $\qhat$, which correspond to the 
expressions in the harmonic-oscillator approximation, can easily be obtained 
by the second moment of Eq.~\eqref{hardqcdresum} up to a UV regulator
$\qhatmu$. In the two limiting 
cases they are thus
\begin{align}
    \label{qhatHOsoft}
    \qhatz(m_D\ll \qhatmu\ll T)=&\als C_R T m_D^2\ln\frac{\qhatmu^2}{m_D^2},\\
    \qhatz(\qhatmu\gg T)= &\frac{4\als^2C_R\zeta(3)T^3}{\pi }
    \Big[2C_A +3 N_f T_F\Big]\ln\frac{\qhatmu^2}{T^2}.\label{qhatHOhard}
\end{align}
Non-logarithmic, $\OO(1)$ constants have been dropped in these expressions. The second one
also features a $\ln(T/m_D)$ contribution.
Numerically, the ratio of the two leading-log coefficients, 
$4\als T^2\zeta(3)\Big[2C_A +3 N_f T_F\Big]/(\pi m_D^2)$, is approximately
0.85 for $N_c=N_f=3$ QCD.

Finally, the $\OO(g)$ correction to $\cc(\lp)$ can be found in \cite{CaronHuot:2008ni}. 
A prescription for connecting it to Eqs.~\eqref{losoft} and \eqref{hardqcd}
without double countings is available in \cite{Ghiglieri:2018ltw}.
For 
$ \lp\ll g^2T$ the screened Coulomb picture is replaced by 
non-perturbative behavior, with $\cc(\lp)\sim \kp^{-3}$ \cite{Schlichting:2021idr}.

\section{Technical details}

\subsection{Details on thermal integrations}
\label{app:integrals}
We start by showing how the thermal part of Eq.~\eqref{eq:kplusint} is derived
under the assumption $\nuuv\gg T\gg \nuir$. The main idea is 
to introduce an intermediate $\epsilon$ regulator, in the spirit of 
dimensional regularization, i.e. 
\begin{align}
    \int^{\nuuv}_{\nuir}\frac{dk^{+}}{k^{+}}\nbe (k^{+})
    &=\lim_{\epsilon\to 0}\left[\int^{\nuuv}_{0}\frac{dk^{+}}{k^{+}}k^{+\epsilon} \nbe (k^{+})
    -\int^{\nuir}_{0}\frac{dk^{+}}{k^{+}}k^{+\epsilon}  \nbe (k^{+})\right]\nonumber\\
    &
    = \lim_{\epsilon\to 0}\left[\int^{\infty}_{0}\frac{dk^{+}}{k^{+}}k^{+\epsilon}  \nbe (k^{+})
    -\int^{\nuir}_{0}\frac{dk^{+}}{k^{+}}k^{+\epsilon}  \left(\frac{T}{k^+}-\frac12\right)\right]
    +\OO\left(\frac{\nuir}{T},e^{-\frac{\nuuv}{T}}\right)\nonumber \\
    &= \lim_{\epsilon\to 0}\left[T^\epsilon \zeta(\epsilon)\Gamma(\epsilon) - 
    \left(\frac{T\nuir^{\epsilon}}{\nuir(\epsilon-1)}-\frac{\nuir^{\epsilon}}{2\epsilon}\right)\right]
    +\OO\left(\frac{\nuir}{T},e^{-\frac{\nuuv}{T}}    \right)\nonumber\\
    &=\frac{T}{\nuir}+\frac12\ln\frac{\nuir e^{\gamma_E}}{2 \pi  T} 
    +\OO\left(\frac{\nuir}{T},e^{-\frac{\nuuv}{T}}    \right)\,. \label{eq:kplusintthermal}
\end{align}
The main advantage is that it allows us to use the known analytically-continued integrations
of the Bose--Einstein distribution in terms of the Riemann $\zeta$ and Euler $\Gamma$ functions.

We now move to employ this method for Eq.~\eqref{eq:BDIMnonzeroTred}. Its thermal part is best 
evaluated by changing the order of the integrals, i.e.
\begin{equation}
    \int_{\tint}^{\frac{\qhatmu^2}{\qhatz}}\frac{d\ftime}{\ftime}
    \int_{\qhatz\ftime^2}^{\qhatmu^2\ftime}\frac{dk^+}{k^+}\nbe(k^+) = 
    \int_{\qhatz\tint^2}^{\qhatmu^2\tint}\frac{dk^+}{k^+}\nbe(k^+)
    \int_{\tint}^{\sqrt{\frac{k^+}{\qhatz}}}\frac{d\ftime}{\ftime}
    +\int^{\frac{\qhatmu^4}{\qhatz}}_{\qhatmu^2\tint}\frac{dk^+}{k^+}\nbe(k^+)
    \int_{\frac{k^+}{\qhatmu^2}}^{\sqrt{\frac{k^+}{\qhatz}}}\frac{d\ftime}{\ftime}.
    \label{eq:BDIMnonzeroTredstart}
\end{equation}
In our $\qhatz\tint^2\ll \omt \ll \qhatmu^2 \tint$ hierarchy, the second integral is exponentially suppressed.
For the first one we can proceed as follows
\begin{align}
    \int_{\qhatz\tint^2}^{\qhatmu^2\tint}\frac{dk^+}{k^+}\nbe(k^+)
    \int_{\tint}^{\sqrt{\frac{k^+}{\qhatz}}}\frac{d\ftime}{\ftime}=
    \frac12\int_{\qhatz\tint^2}^{\qhatmu^2\tint}\frac{dk^+}{k^+}\nbe(k^+)
    \ln\frac{k^+}{\qhatz\tint^2}.
\end{align}
We can then exploit that 
\begin{equation}
    \label{polygamma}
    \int_{0}^{\infty}\frac{dk^+}{k^+}k^{+\epsilon}\nbe(k^+)\ln\frac{k^+}{T}=
    T^{\epsilon} \Gamma (\epsilon) \left(\zeta '(\epsilon)+\zeta (\epsilon) \psi(\epsilon)\right),
\end{equation}
where $\psi(x)$ is the digamma function. This, together with Eq.~\eqref{eq:kplusintthermal}, leads to 
\begin{align}
    \int_{\qhatz\tint^2}^{\qhatmu^2\tint}\frac{dk^+}{k^+}\nbe(k^+)
    \int_{\tint}^{\sqrt{\frac{k^+}{\qhatz}}}\frac{d\ftime}{\ftime}
   =&\frac{T}{2\qhatz\tint^2}+\frac18\left(- \ln^2\frac{\omt}{\qhat\tint^2}
  + \gamma_E^2 - \frac{\pi^2}{4}  +2\gamma_1\right)\nonumber\\
  &+\OO\left(\frac{\qhatz\tint^2}{T},
  e^{-\frac{\qhatmu^2\tint}{T}}\right),
\end{align}
which in turn can be added to the straightforward vacuum part to give
\begin{align}
    \delta \qhat(\qhatmu)^{\mathrm{few}}=
    \frac{\als C_R}{2\pi}\qhatz\bigg\{\frac{2T}{\qhatz\tint^2}+
    \ln^2\frac{\mu ^2}{\qhatz \tint} 
    -\frac{1}{2}\ln^2\frac{\omt}{\qhatz\tint^2} + \frac{\gamma_E^2}{2} - \frac{\pi^2}{8}  +\gamma_1
    +\ldots
  \bigg\},
\end{align}
whose purely double-logarithmic contribution we have anticipated in Eq.~\eqref{eq:BDIMresultnonzeroT}. 
The dots stand for the suppressed terms. Eq.~\eqref{semievaltaufinal} can be obtained using the 
same techniques.

Let us turn to Eq.~\eqref{semievaltaufinalnon}. The starting point is 
\begin{align}
    \mathcal{I}\equiv
    \int_{\kpir}^{\qhatmu^2\tint}\frac{dk^+}{k^+}
    \left(\frac12+\nbe(k^+)\right) \int_{k^+/\qhatmu^2}^{\tint}\frac{d\ftime}{\ftime}
    \left[-\frac{1}{4\ftime^{2}}\ln\left(1+4m_D^2\ftime^2\right)-
    m_D^2\ln\left(1+\frac{1}{4m_D^2\ftime^2}\right)\right].
\end{align}
The $\ftime$ integration yields
\begin{align}
    \mathcal{I}=
    \int_{\kpir}^{\qhatmu^2\tint}\frac{dk^+}{k^+}
    \left(\frac12+\nbe(k^+)\right) \bigg[&\frac{m_D^2}{2} \mathrm{Li}_2
    \left(-\frac{\qhatmu^4}{4k^{+2}m_D^2}\right)
    -\frac{m_D^2}{2}\ln\left(1+\frac{\qhatmu^4}{4k^{+2}m_D^2}\right) \nonumber \\
    &-\frac{\qhatmu^4}{8k^{+2}}\ln\left(1+\frac{4k^{+2}m_D^2}{\qhatmu^4}\right)
    +\OO\left(\frac{1}{\tint^2}\right) \bigg],
\end{align}
where $\mathrm{Li}_2$ is the dilogarithm and we have expanded for $m_D\tint\gg 1$, recalling that $1/gT\ll \tint\ll 1/g^2T$.
The vacuum ($1/2$) part needs to be integrated as is, without further expansions: while close to 
the IR boundary we could exploit that $\qhatmu^2/(2k^+m_D)\gg 1$, that would not be true 
close to the UV boundary, where instead $\qhatmu^2/(2k^+m_D)\ll 1$. The integral is however
not problematic and yields
\begin{align}
    \mathcal{I}_\mathrm{vac}=&
    -\frac{1}{6} m_D^2 \bigg[\ln ^3\left(\frac{\qhatmu ^2}{2 \kpir
    m_D}\right)+\frac{3}{2} \ln ^2\left(\frac{\qhatmu ^2}{2 \kpir
    m_D}\right)+\frac{1}{4} \left(6+\pi ^2\right) \ln
    \left(\frac{\qhatmu ^2}{2 \kpir m_D}\right)+\frac{1}{8} \left(6+\pi
    ^2\right)\bigg]\nonumber\\
    &+\OO\left(\frac{1}{\tint^2},(\kpir)^2\right),\label{Ivacfinal}
\end{align}
where we have again expanded in the cutoffs. For the thermal part we can on 
the other hand expand for $\qhatmu^2/(2k^+m_D)\gg 1$, as the effective UV cutoff
introduced by Boltzmann suppression makes the region where $\qhatmu^2/(2k^+m_D)\lesssim 1$ exponentially
suppressed. Hence
\begin{align}
    \mathcal{I}_{T}=-m_D^2
    \int_{\kpir}^{\infty}\frac{dk^+}{k^+}
   \nbe(k^+)\bigg[&\ln^2\frac{\qhatmu^2}{2k^{+}m_D}
    +\ln\frac{\qhatmu^2}{2k^{+}m_D}
    +\frac{1}{2}\left(1+\frac{\pi^2}{6}\right)\bigg]+\ldots,
\end{align}
where the dots stand for higher-order terms in the expansions in the cutoffs and for the exponentially
suppressed term arising from approximating the UV cutoff to infinity. This integral can be done with 
Eqs.~\eqref{eq:kplusintthermal}, \eqref{polygamma} and
\begin{equation}
    \label{polygamma2}
    \int_{0}^{\infty}\frac{dk^+}{k^+}k^{+\epsilon}\nbe(k^+)\ln^2\frac{k^+}{T}=
    T^{\epsilon} \Gamma (\epsilon) \left[\zeta ''(\epsilon)+2 \psi(\epsilon)
   \zeta '(\epsilon)+\zeta (\epsilon) \left(\psi (\epsilon)^2+\psi
   ^{(1)}(\epsilon)\right)\right],
\end{equation}
where $\psi^{(1)}$ is the polygamma function of order 1. Using these three integrals we find
\begin{align}
    \mathcal{I}_T=&-\frac{m_D^2 T \left( \ln^2 \left(\frac{\qhatmu ^2}{2 \kpir
    m_D}\right)- \ln \left(\frac{\qhatmu ^2}{2 \kpir
    m_D}\right)+\frac{\pi ^2}{12}+\frac32\right)}{ \kpir}\nonumber\\
    &+\frac{m_D^2}{6}\bigg\{
           \ln ^3\left(\frac{\qhatmu ^2}{2 \kpir
        m_D}\right)- \ln ^3\left(\frac{\qhatmu ^2}{2 m_D
        \omt}\right)
        +
        \frac32\ln^2\left(\frac{\qhatmu ^2}{2 \kpir m_D}\right)-\frac32 \ln ^2\left(\frac{\qhatmu
        ^2}{2 m_D \omt}\right)\nonumber \\
        &+
        \frac14\left(6+\pi ^2\right) \ln \left(\frac{\qhatmu ^2}{2 \kpir
        m_D}\right)+\ln \left(\frac{\qhatmu ^2}{2 m_D
        \omt}\right)\left[6 \gamma _1 -\pi ^2 +3 \gamma_E ^2 -\frac32 \right]\nonumber \\
        &
        -6 \gamma_E  \gamma _1-3 \left(-
        \gamma _1+ \gamma _2+\frac{\pi ^2}{8}\right)+\frac12 (3-4 \gamma_E ) \gamma_E ^2    
    \bigg\}+\ldots\,.\label{Itherm}
\end{align}
Upon summing Eqs.~\eqref{Ivacfinal} and \eqref{Itherm} we see that 
all logarithmic dependence on $\kpir$ vanishes, yielding
\begin{align}
    \mathcal{I}=&-\frac{m_D^2 T \left( \ln^2 \left(\frac{\qhatmu ^2}{2 \kpir
    m_D}\right)- \ln \left(\frac{\qhatmu ^2}{2 \kpir
    m_D}\right)+\frac{\pi ^2}{12}+\frac32\right)}{ \kpir}\nonumber\\
    &+\frac{m_D^2}{6}\bigg\{
        - \ln ^3\left(\frac{\qhatmu ^2}{2 m_D
        \omt}\right)
       -\frac32 \ln ^2\left(\frac{\qhatmu^2}{2 m_D \omt}\right)
        +\ln \left(\frac{\qhatmu ^2}{2 m_D
        \omt}\right)\left[6 \gamma _1 -\pi ^2 +3 \gamma_E ^2 -\frac32 \right]\nonumber \\
        &
        -6 \gamma_E  \gamma _1-3 \left(-
        \gamma _1+ \gamma _2\right)+\frac12 (3-4 \gamma_E ) \gamma_E ^2    
        -\frac{1}{4} \left(3+2\pi
        ^2\right)\bigg\}+\ldots\,.\label{Ifinal}
\end{align}
Upon reinstating the proper prefactor we have
\begin{equation}
    \delta\qhat_\mathrm{semi}^{l^--\mathrm{dep}}=\frac{2\als^2 C_R C_A T}{\pi} \mathcal{I},
    \label{semievaltaufinalnontot}
\end{equation}
whose highest logarithmic term we anticipated in Eq.~\eqref{semievaltaufinalnon}.

\subsection{Hard semi-collinear subtraction}
\label{app:AX}
In this Appendix, we will show how the second term in Eq.~\eqref{defqhatde} 
is automatically taken into account if we add the hard contribution 
to $\hat{q}$ from \cite{Arnold:2008vd}. Indeed, that paper 
computes the contribution to $\qhat$ for $\sqrt{ET}\gg \kp\gg gT$ at leading order,
i.e. through elastic Coulomb scatterings with the light quarks and gluons of the medium.
The  incoming and outgoing momenta for such scatterers are named
$P_2$ and $P_2-Q$ there. We can then identify $P_2$ with our $L$ and $Q$ with $-K$.\footnote{
In the language of Eq.~\eqref{hardqcd}, which reproduces the hard contribution 
to $\cc(\lp)$, we should identify $q$ there with $l$ here and $l$ there with $k$ here.
}
In order to obtain the contribution to $\cc(\kp)$ and thence $\qhat$, \cite{Arnold:2008vd}
integrates over all values  and orientations of $p_2$. In so doing, it 
ends up including the slice where $p_2\sim gT$, $\kp\sim\sqrt{g}T$, which 
is precisely the semi-collinear scaling we investigated in Sec.~\ref{sec:semi}.
As explained there, we need to subtract this limit of \cite{Arnold:2008vd}, so 
as to avoid double countings. We thus proceed to its determination.

Our starting point is Eq.~(3.8) of \cite{Arnold:2008vd}, where we have specialized to the case 
of scattering off a soft gluon ($\nbe(l^0)\approx T/l^0$) 
and undone a couple of auxiliary integrals (see (3.7) there), 
as well as applied the dictionary just described. We then have
\begin{align}
    \delta\mathcal{C}(k_{\perp})^{\text{hard}}&=
    \frac{2g^{4}C_RC_{A}}{k_{\perp}^{4}}\int\frac{dk^{0}}{2\pi}\int\frac{dk^{z}}{2\pi}\int\frac{d^{4}L}
    {(2\pi)^{4}}(l^{0}-l^{z})^{2}\frac{T}{l^{0}}(1+\nbe (l^0+k^0))\nonumber
    \\&\times 2\pi\delta(k^{0}-k^{z})2\pi\delta((L+K)^{2})2\theta(l^{0}+k^{0})2
    \theta(l^{0})2\pi\delta(L^{2}).
\end{align}
For a soft gluon $L\sim gT$  we can simplify the 
expression above as
\begin{align}
    \delta\mathcal{C}(k_{\perp})^{\text{hard}}&=
    \frac{2g^{4}C_RC_{A}}{k_{\perp}^{4}}\int\frac{dk^{+}}{2\pi}\int\frac{d^{4}L}
    {(2\pi)^{4}}\frac{Tl^{-2}}{l^++l^-/2}(1+\nbe (k^+))\nonumber
    \\&\times 2\pi\delta(2k^+l^--\kp^2)2\theta(k^{+})2
    \theta(l^++l^-/2)2\pi\delta(L^{2}).
\end{align}
We can set up the $\delta$ function to fix $l^+$, in 
analogy to what has been done in Sec.~\ref{sec:semi}, i. e.
\begin{align}
    \delta\mathcal{C}(k_{\perp})^{\text{hard}}&=
    \frac{2g^{4}C_RC_{A}}{k_{\perp}^{4}}\int\frac{dk^{+}}{2\pi}\int\frac{d^{4}L}
    {(2\pi)^{4}}\frac{Tl^{-2}}{l^++l^-/2}(1+\nbe (k^+))\frac{\theta(k^{+})}{k^+}\nonumber
    \\&\times 2\pi\delta\left(l^--\frac{\kp^2}{2k^+}\right)
    \frac{\theta(l^++l^-/2)}{\abs{l^-}}2\pi\delta\left(l^+-\frac{\lp^2}{2l^-}\right).
\end{align}
This finally yields
\begin{equation}
    \label{semihardfinal}
    \delta\mathcal{C}(k_{\perp})^{\text{hard}} = \frac{g^{2}C_R}{\pi k_{\perp}^{4}}
    \int\frac{dk^{+}}{k^+}(1+\nbe (k^+))\theta(k^{+})\int\frac{d^2\lp}
    {(2\pi)^2}\frac{g^2 C_A T\;2l^{-2}}{\lp^2+l^{-2}}\bigg\vert_{l^-=\frac{\kp^2}{2k^+}}.
\end{equation}

It is encouraging to see the resemblance of this result to that of the second term of Eq.~\eqref{defqhatde} 
when plugged in Eq.~\eqref{resultfromsemi2}. 
However, to get the two results to match exactly, we need to 
compute the same result with same integral with instead $L\to L-K$ and $L+K\to L$ keeping 
$L$ soft, that is, a soft outgoing gluon scatterer. 
This second integral will give the negative $k^+$ contribution, and the integrals' sum will indeed yield 
\begin{equation}
    \label{resultfromsemi2app}
    \delta\cc(\kp)_\mathrm{semi}^{\text{hard}}= \frac{g^2 C_R}{\pi\kp^4}\int\frac{dk^+}{k^+}(1+\nbe(k^+))
    \qhat_{\text{\cite{Arnold:2008vd}}}\left(\homu;\frac{\kp^2}{2k^+}\right).
\end{equation}

\subsection{$T/\kpir$ terms from the soft, classical calculation}
\label{app_arc}

As we mentioned in Sec.~\ref{sub:classical}, this appendix is devoted to 
showing how the results of \cite{Ghiglieri:2015ala} can be used 
to explicitly derive the $T/\kpir$ terms that arise on the 
$\mathcal{A}_R$ contour of Fig.~\ref{fig:matsubara} in the 
reduction to EQCD derived in~\cite{CaronHuot:2008ni}.
For reasons that shall soon become clear, 
let us start from Eq.~(F.49) of \cite{Ghiglieri:2015ala}. It reads
\begin{eqnarray}
    \label{almostfinalsoft}
      \delta\qhat_\mathrm{L}\Big\vert_{\mathrm{loop}} \!\!\!
      -\delta\qhat_\mathrm{L}\Big\vert_{\mathrm{subtr.}}^\mathrm{diff} \!\! &=&
    4g^4C_R C_A T
    \! \int_{\mathcal{C}_R} \! \frac{dq^+}{2\pi}
    \! \int \! \frac{d^2\qp}{(2\pi)^2}
    \! \int \! \frac{d^4K}{(2\pi)^4}
    \\ && \hspace{-1cm}\times  \nonumber
    \bigg\{
    \frac{G^{--}_{rr}(K)\pi\delta(k^-)}{\qp^2+m_\infty^2}
    \left(\frac{\qp^2}{\qp^2+m_\infty^2} -\frac{\qp^2+\bqp\cdot\bkp}{(\bqp+\bkp)^2+m_\infty^2}\right)
    +\OO\left(\frac{1}{q^+}\right)\bigg\} +\mathcal{C}_A,
    \end{eqnarray}
where we have dropped subleading terms on the $\mathcal{C}_R$ and $\mathcal{C}_A$ arcs, which are defined below
Eq.~(3.19) in \cite{Ghiglieri:2015ala}.  $G^{--}(K)$ denotes the Coulomb-gauge 
HTL propagator, as per App.~\ref{sec_conv}.
We can now symmetrize the expression in round brackets, leading to
\begin{eqnarray}
    \label{almostfinalsoftsym}
      \delta\qhat_\mathrm{L}\Big\vert_{\mathrm{loop}} \!\!\!
      -\delta\qhat_\mathrm{L}\Big\vert_{\mathrm{subtr.}}^\mathrm{diff} \!\! &=&
    2g^4C_R C_A T
    \! \int_{\mathcal{C}_R} \! \frac{dq^+}{2\pi}
    \! \int \! \frac{d^2\qp}{(2\pi)^2}
    \! \int \! \frac{d^4K}{(2\pi)^4}
    \\ &&\hspace{-1cm} \times  \nonumber
    \bigg\{
    G^{--}_{rr}(K)\pi\delta(k^-)
    \left(\frac{\bqp}{\qp^2+m_\infty^2} -\frac{\bqp+\bkp}{(\bqp+\bkp)^2+m_\infty^2}\right)^2
    +\OO\left(\frac{1}{q^+}\right)\bigg\} +\mathcal{C}_A.
    \end{eqnarray}
We can now observe that
this is an expression for the longitudinal momentum diffusion 
    $\qhat_L\equiv \langle q_z^2\rangle/\lmed$. It can be translated
    to its transverse counterpart by multiplying  by $\qp^2/q^{+2}$ under the integral sign.
We can further translate to our momentum coordinates with the dictionary 
    mentioned in Sec.~\ref{sec:semi}, i.e. $Q_{\text{\cite{Ghiglieri:2015ala}}}\to K$, 
   $K_{\text{\cite{Ghiglieri:2015ala}}}\to L$. By comparing the definition of the 
    arcs there with ours, as per Fig.~\ref{fig:matsubara}, we can also 
    translate $\mathcal{C}_R$ to $\mathcal{A}_R$.\footnote{$\mathcal{C}_R$ is 
    clockwise, whereas $\mathcal{A}_R$ is counterclockwise. But what we want is only the
    red, horizontal part of the contour  in Fig.~\ref{fig:matsubara},
    so we should be subtracting $\mathcal{A}_R$, thus equating it with $\mathcal{C}_R$.
    Hence in a bit of a misnomer the $\mathcal{A}_R$ contribution in this appendix
    is minus the contour in Fig.~\ref{fig:matsubara}.}
This yields
\begin{eqnarray}
    \label{qhatarc}
      \delta\qhat\Big\vert_{\mathrm{soft}}^\mathrm{arcs} \!\! &=&
    2g^4C_R C_A T
    \! \int_{\mathcal{A}_R} \! \frac{dk^+}{2\pi}
    \! \int \! \frac{d^2\kp}{(2\pi)^2}
    \! \int \! \frac{d^4L}{(2\pi)^4}
    \\ && \times  \nonumber
    \bigg\{
    \frac{\kp^2}{k^{+2}}G^{--}_{rr}(L)\pi\delta(l^-)
    \left(\frac{\bkp}{\kp^2+m_\infty^2} -\frac{\bkp+\blp}{(\bkp+\blp)^2+m_\infty^2}\right)^2
   +\OO\left(\frac{1}{k^{+3}}\right)\bigg\} +\mathcal{A}_A.
    \end{eqnarray}
We can now perform some of the $L$ integrations using the reduction to EQCD 
discussed around Eq.~\eqref{treeAGZ}, finding
\begin{eqnarray}
    \label{qhatarcint}
      \delta\qhat\Big\vert_{\mathrm{soft}}^\mathrm{arcs} \!\! &=&
    g^4C_R C_A T^2
    \! \int_{\mathcal{A}_R} \! \frac{dk^+}{2\pi}
    \! \int \! \frac{d^2\kp}{(2\pi)^2}
    \! \int \! \frac{d^2\lp}{(2\pi)^2}
    \\ && \times  \nonumber
    \bigg\{
    \frac{\kp^2}{k^{+2}}\frac{m_D^2}{\lp^2(\lp^2+m_D^2)}
    \left(\frac{\bkp}{\kp^2+m_\infty^2} -\frac{\bkp+\blp}{(\bkp+\blp)^2+m_\infty^2}\right)^2
    +\OO\left(\frac{1}{k^{+3}}\right)\bigg\} +\mathcal{A}_A.
    \end{eqnarray}
If we take the $\lp,m_D\ll\kp$ approximation, consistently
with our treatment across the $\kpir$ boundary, we find
\begin{eqnarray}
    \label{qhatarcintsemi}
      \delta\qhat\Big\vert_{\mathrm{soft}}^\mathrm{arcs} \!\! &=&
    g^4C_R C_A T^2
    \! \int_{\mathcal{A}_R} \! \frac{dk^+}{2\pi}
    \! \int \! \frac{d^2\kp}{(2\pi)^2}
    \! \int \! \frac{d^2\lp}{(2\pi)^2}
    \frac{1}{k^{+2}\kp^2}\frac{m_D^2}{\lp^2+m_D^2}
 +\mathcal{A}_A.
    \end{eqnarray}
If we use the harmonic-oscillator approximation to regulate the $\lp$ integration,
consistently with what we did in Sec.~\ref{sec:semi}, we find
\begin{eqnarray}
    \label{qhatarcintsemiHO}
      \delta\qhat\Big\vert_{\mathrm{soft}}^\mathrm{arcs\,HO} \!\! &=&
    \frac{g^4C_R C_A T^2m_D^2}{4\pi}
    \! \int_{\mathcal{A}_R} \! \frac{dk^+}{2\pi}
    \! \int \! \frac{d^2\kp}{(2\pi)^2}
    \frac{1}{k^{+2}\kp^2}\ln\left(\frac{\homu^2}{m_D^2}\right)
 +\mathcal{A}_A.
    \end{eqnarray}
We can then rewrite the $\kp$ integration as a $\ftime$ one, with the same 
boundaries of Eq.~\eqref{semievaltau}, i.e.
\begin{eqnarray}
    \label{qhatarcintsemiHOlm}
      \delta\qhat\Big\vert_{\mathrm{soft}}^\mathrm{arcs\,HO} \!\! &=&
    \frac{g^4C_R C_A T^2m_D^2}{4\pi}
    \! \int_{\mathcal{A}_R} \! \frac{dk^+}{2\pi}
    \! \int_{k^+/\qhatmu^2}^{\tint} \!
    \frac{d\ftime}{4\pi}
    \frac{1}{k^{+2}\ftime}\ln\left(\frac{\homu^2}{m_D^2}\right)
 +\mathcal{A}_A.
    \end{eqnarray}
The contribution on the arc can be carried out as follows, understanding the 
arc to be at constant $\vert k^+\vert = \kpir$. Changing variables to $k^+=\kpir e^{it}$ then gives
\begin{equation}
    \int_{\mathcal{A}_R} \! \frac{dk^+}{2\pi}\frac{1}{k^{+2}}
    \ln\left(\frac{\qhatmu^2\tint}{k^+}\right)=
    \frac{i}{2\pi\kpir} \int_\pi^{0} dt e^{-i t}
    \bigg[\ln\left(\frac{\qhatmu^2\tint}{\kpir}\right)
    -it\bigg]= \frac{-1}{\pi\kpir}\bigg[\ln\left(\frac{\qhatmu^2\tint}{\kpir}\right)
    -1-i\frac\pi2\bigg].
\end{equation}
The imaginary part cancels against an opposite one from $\mathcal{C}_A$. We then have
\begin{align}
      \delta\qhat\Big\vert_{\mathrm{soft}}^\mathrm{arcs\,HO} \!\! =&
    2\frac{g^4C_R C_A T^2m_D^2}{16\pi^2}
    \frac{-1}{\pi\kpir}\bigg[\ln\left(\frac{\qhatmu^2\tint}{\kpir}\right)
    -1\bigg]
   \ln\left(\frac{\homu^2}{m_D^2}\right)\nonumber \\
   =&-2\frac{\als C_R  T\qhatz(\homu)}{\pi\kpir}
  \bigg[\ln\left(\frac{\qhatmu^2\tint}{\kpir}\right)
  -1\bigg]
  .\label{qhatarcintsemiHOalmostfinal}
    \end{align}
This is precisely opposite to the
$T/\kpir$-proportional part of Eq.~\eqref{semievaltaufinal}, as 
we set out to show. As a final remark, the cancellation of the $T/\kpir$ terms associated 
with the $\ftime>\tint$ contribution requires a different calculation, which
we do not show.

\section{Diagrammatic evaluation of the semi-collinear and virtual processes}
\label{app:diagrams}
\begin{figure}[ht]
    \begin{center}
        \includegraphics[width=12cm]{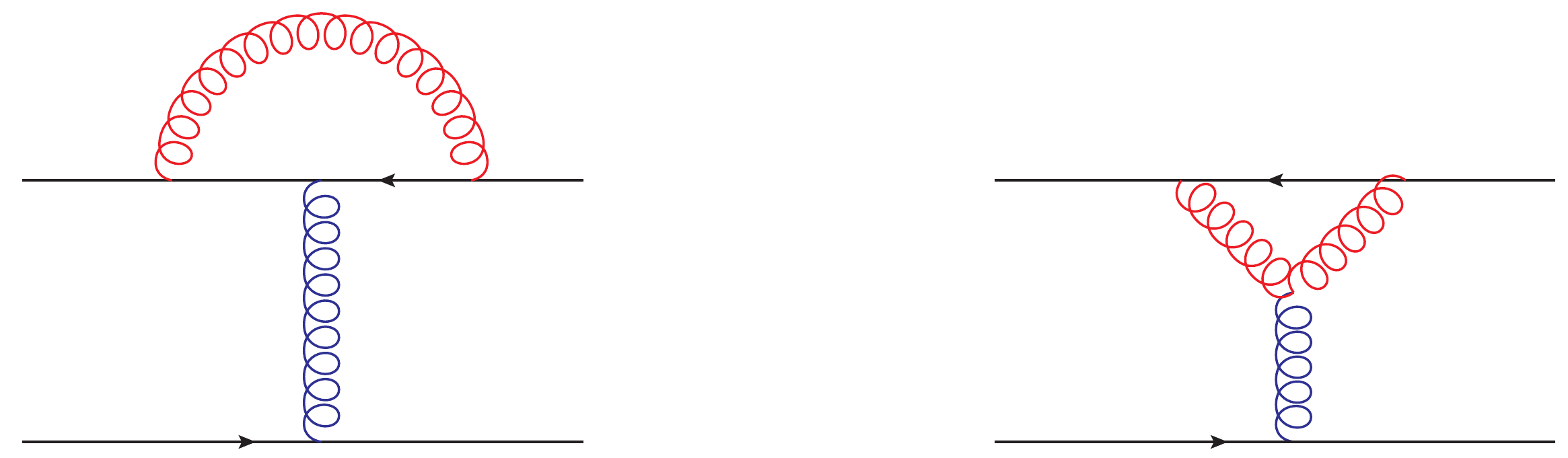}
    \end{center}
    \caption{ Diagrams whose cuts correspond to virtual processes.}
    \label{fig:virtual}
\end{figure}
In this appendix we provide a sketch of the diagrammatic evaluation of the diagrams in 
Fig.~\ref{fig:real} in the semi-collinear scaling. In addition, we shall also
consider the \emph{virtual} counterpart, as depicted in Fig.~\ref{fig:virtual}. The 
name labels the fact that the coft gluon is  virtual, and transverse momentum 
exchange happens exclusively via the soft interactions with the medium. The only 
nonvanishing cut in these diagrams goes through the soft gluon's HTL, and corresponds
to the interference of the leading-order soft scattering with the medium with its 
coft-gluon-loop virtual correction.

\begin{figure}[ht]
    \begin{center}
        \includegraphics[width=12cm]{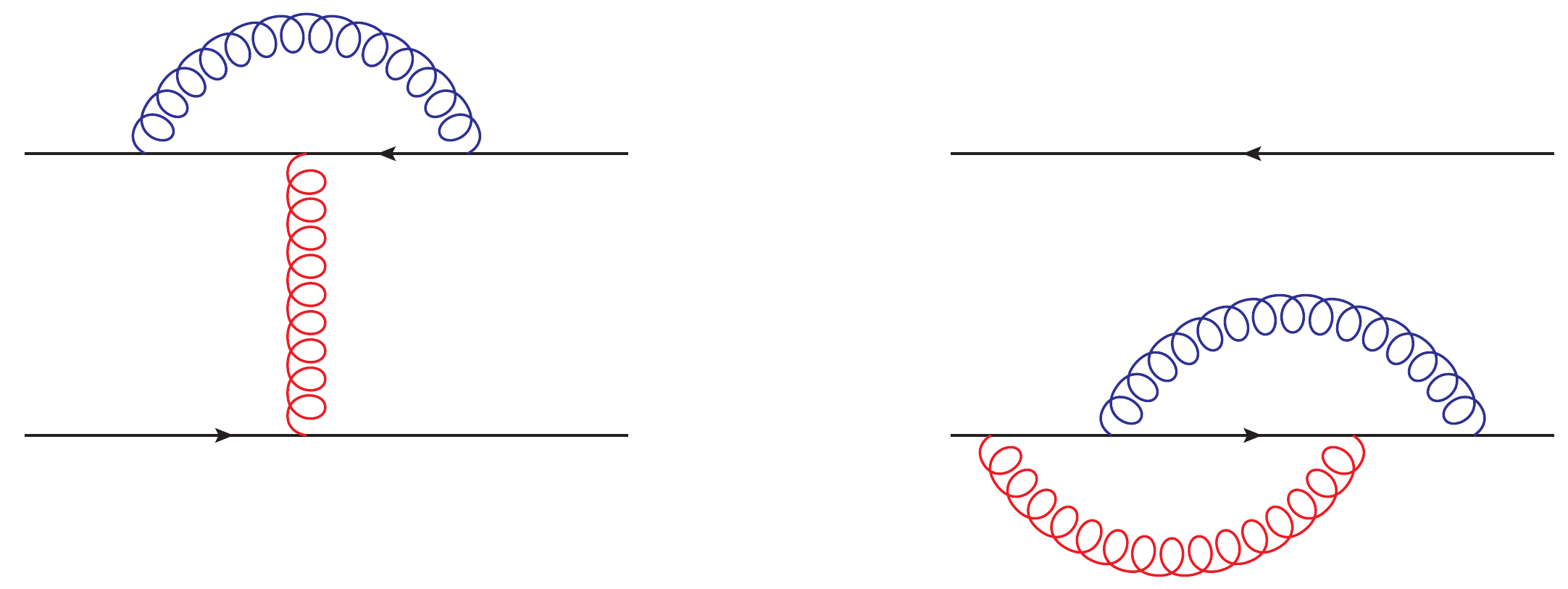}
    \end{center}
    \caption{Two type of diagrams that do not contribute to $\cc(\kp)$, see main text.}
    \label{fig:vanish}
\end{figure}
The attentive reader will have noticed that the diagrams of Figs.~\ref{fig:real}
and \ref{fig:virtual} do not form the complete set at that order. Missing diagrams are part 
of two classes. Fig.~\ref{fig:vanish} depicts an example for each. On the left we have 
diagrams that vanish because their only cut goes through a single coft gluon. 
Coft gluons have vanishing spectral weight at $k^-=0$, which is imposed
by the $x^+$ integrations at large $\lmed$. On the right instead we have  
$\xp$-independent diagrams:
they thus contribute to the constant part of $\cc(\xp)$ which is related to probability 
conservation. Hence, by looking at Eqs.~\eqref{eq:cdef} and \eqref{eq:cpos} 
we can get the contributions 
to $\cc(\kp)$ by only considering non-vanishing, $\xp$-dependent diagrams and
 undoing the overall Fourier transform.

\subsection{Real processes}
\label{sub:real}
We start from the  two diagrams on the top line in Fig.~\ref{fig:real}. Their combination 
contains the $C_R^2$-proportional 
cross term of the one-coft-gluon exchange with the one-soft-gluon exchange, which is part 
of the \emph{exponentiation} of the tree-level contribution. As the one-coft-gluon exchange
vanishes for the kinematical reasons just mentioned, only the $C_RC_A$-proportional, non-abelian
piece of the top right diagram survives. 
It is given by 
\begin{align}
    \delta\mathcal{C}(k_{\perp})_{\mathrm{II+X}}&=
  C_{R}C_{A}g^{4} \int\frac{dk^{+}dk^-}{(2\pi)^2}\int\frac{d^{4}L}{(2\pi)^{4}}
  \frac{1}{(l^{-}-i\epsilon)^{2}}G_{>}^{--}(K+L)G_{>}^{--}(-L)2\pi\delta(k^-). 
\end{align}
II and X reflect the topology of the two diagrams, whereas the propagators are both 
Wightman because of the path-ordering of the fields in the Wilson loop. However, the
$L$ propagator is to be understood as soft, and thus HTL-resummed, whereas the $K+L$, being
coft, can be taken as bare. A factor of 2 has been added to account
for the inverted scaling.  Coulomb gauge is implied here and in the rest of the calculation.
The $\delta(k^-)$ arises from the $x^+$ integrations at large $\lmed$.
We do not proceed further with the evaluation, as we prefer to combine all diagrams before. 

We now proceed to the self-energy diagram on the bottom left in Fig.~\ref{fig:real}.
Its contribution reads
\begin{align}
    &\delta\mathcal{C}(k_{\perp})_{\mathrm{self}}=g^{4}C_{R}C_{A}
    \int\frac{dk^{+}}{2\pi}\int \frac{d^{4}L}{(2\pi)^{4}}G_{R}^{-\rho}(K)\Big(g_{\gamma\sigma}(2L{+}K)_{\rho}
    -g_{\sigma\rho}(2K{+}L)_{\gamma}+g_{\rho\gamma}(K{-}L)_{\sigma}\Big)\nonumber
    \\&\times G_{>}^{\sigma\delta}(K+L)G_{>}^{\gamma\alpha}(-L)
    \Big(g_{\delta\beta}(2K{+}L)_{\alpha}-g_{\alpha\delta}(2L{+}K)_{\beta}-g_{\alpha\beta}(K{-}L)_{\delta}\Big) 
    G_{A}^{\beta-}(K)\Big\vert_{k^{-}=0}.
    \label{eq:aftercut}
\end{align}
Cuts going through the $K$ propagators are again vanishing, hence their retarded/advanced
assignments only, which is consistent with the cutting rules for Wightman functions 
\cite{Caron-Huot:2007zhp,Ghiglieri:2020dpq}.

Finally, the Y-shaped diagram on the bottom right  of Fig.~\ref{fig:real}, together with
its symmetry-related counterparts, yields
\begin{align}
    \delta\mathcal{C}(x_{\perp})_{\mathrm{Y}}=&2g^{4}C_{R}C_{A}\int 
    \frac{dk^+d^2\kp}{(2\pi)^{3}}\int \frac{d^{4}L}{(2\pi)^{4}}
     \frac{i}{l^{-}+i\epsilon}e^{i\bkp\cdot \bxp }G_{>}^{-\delta}(K+L)G_{>}^{-\alpha}(-L)\nonumber
    \\&\hspace{1.2cm}\times\Big(g_{\alpha\delta}(2L+K)_{\beta}-g_{\delta\beta}(2K+L)_{\alpha}
    +g_{\beta\alpha}(K-L)_{\delta}\Big)G_{F}^{\beta-}(K)\Big\vert_{k^{-}=0}.
\end{align}
At this point, it is useful to write the Feynman propagator as 
\begin{equation}
    \label{eq:feynman}
    G_{F}(K)=\frac{1}{2}\Big(G_{R}(K)+G_{A}(K)\Big)+G_{rr}(K).
\end{equation}    
Since $G_{rr}$ is the average of the bare cut propagators and we have already set $k^{-}=0$, 
it vanishes. For the same reason, we have not considered the other
real-time assignments. 
That leaves us with the average of the retarded and advanced bare Green's functions, 
which when evaluated at $k^{-}=0$ are the same. We can then
extract, as was done with the previous diagrams
\begin{align}
    \delta\mathcal{C}(k_{\perp})_{\mathrm{Y}}=&-2g^{4}C_{R}C_{A}\int \frac{dk^{+}}{2\pi}
    \int \frac{d^{4}L}{(2\pi)^{4}}
    \frac{i}{l^{-}+i\epsilon}G_{>}^{-\delta}(K+L)G_{>}^{-\alpha}(-L)
    \nonumber
   \\&\hspace{1.2cm}\times\Big(g_{\alpha\delta}(2L+K)_{\beta}-g_{\delta\beta}(2K+L)_{\alpha}+
   g_{\beta\alpha}(K-L)_{\delta}\Big)G_{R}^{\beta-}(K)\Big\vert_{k^{-}=0}.
\end{align}
We can then get the semi-collinear rate by summing the contribution from the 
three diagrams, i.e.
\begin{equation}
    \label{defsemidiag}
    \delta\mathcal{C}(k_{\perp})_\mathrm{semi} = \delta\mathcal{C}(k_{\perp})_{\mathrm{II+X}}
    +\delta\mathcal{C}(k_{\perp})_{\mathrm{self}}+ \delta\mathcal{C}(k_{\perp})_{\mathrm{Y}}.
\end{equation}
Upon taking care of Lorentz algebra and expanding the resulting expression 
for the scaling $L\sim gT$, $k^+\sim T,k_\perp\sim \sqrt{g}T$ 
we obtain
\begin{align}
    \delta\mathcal{C}(k_{\perp})_\mathrm{semi}
    =4g^{4}C_{R}C_{A}\int\frac{dk^{+}}{2\pi}\int\frac{d^{4}L}{(2\pi)^{4}}\frac{G^{T}_{>}(K+L)}
    {k_{\perp}^{4}}\Bigg(&l_{\perp}^{2}G^{--}_{rr}(L)\nonumber
    \\& 
    +2G^{T}_{rr}(L)\Big({l^{-}}^{2}-\frac{l^{+}l^{-}l_{\perp}^{2}}{l^2}\Big)\Bigg)_{k^-=0}.
\end{align}
Upon inserting the form of the bare propagator, expanded under this scaling, we find
\begin{align}
    \delta\mathcal{C}(k_{\perp})_\mathrm{semi}=&2g^{4}C_{R}C_{A}\int\frac{dk^{+}}{2\pi}\frac{1+\nbe (k^+)}{k_{\perp}^{4}k^{+}}\int\frac{d^{4}L}{(2\pi)^{3}}
    \delta\left(l^{-}-\frac{k_{\perp}^{2}}{2k^{+}}\right)\Bigg(l_{\perp}^{2}G^{--}_{rr}(L)\nonumber
    \\&\hspace{5cm}+2G^{T}_{rr}(L)\Big({l^{-}}^{2}-\frac{l^{+}l^{-}l_{\perp}^{2}}{l^2}\Big)\Bigg).
\end{align}
The $l^+$ integration can be addressed using light-cone analyticity, as per 
\cite{Ghiglieri:2013gia}, leading to Eq.~\eqref{resultfromsemi2}.

\subsection{Virtual processes}
\label{sub:virtual}

We start from the first diagram in Fig.~\ref{fig:virtual}. As was the case for the real II and 
X diagrams, the $C_R^2$-proportional, abelian part of this diagram vanishes 
with the counterparts, not shown in the figure, where the coft gluon does not straddle
the soft one. Accounting for this and for the symmetric diagrams with the coft gluon
attached to the other Wilson line we have
\begin{equation}
    \delta\mathcal{C}(l_{\perp})^\mathrm{virt}_{\widehat{\mathrm{T}}}=-g^{4}C_{R}C_{A}\int\frac{d^{4}K}{(2\pi)^{4}}
    \int\frac{dl^{+}}{2\pi}\frac{1}{(k^{-}+i\epsilon)^{2}}
G_{>}^{--}(L)G_{rr}^{--}(K)\Big\vert_{l^{-}=0},
\end{equation}
where we have labeled this diagram $\widehat{\mathrm{T}}$ following its topology. We note that 
the Wilson line integration forces $l^-=0$, differently from the real processes.
In this case too
we defer the evaluation of this expression until after the second diagram has been evaluated. 
Its contribution, together with that of its symmetric counterpart with two vertices on the bottom Wilson line, 
is\footnote{At first glance, it is not clear why we cannot have any other real-time 
assignments for this diagram. Indeed, one could imagine the
assignment $G_{F}(L)G_{>}(K+L)G_{>}(-L)$. 
We have checked that this leaves us with an expression that, at first order in the collinear expansion,
is odd in $k^+$ and thus integrates to zero.}
\begin{align}
    \delta\mathcal{C}(x_{\perp})^\mathrm{virt}_{{\mathrm{Y}}}=&\frac{g^{4}C_{R}C_{A}}{2}
    \int\frac{d^{4}K}{(2\pi)^{4}}\int\frac{d^4L}{(2\pi)^{4}}\frac{-i}{k^{-}-i\epsilon}
    e^{i\blp\cdot \bxp} G_{>}^{\beta-}(L)2\pi\delta(l^-)\nonumber
    \\&\hspace{-2cm}\times\big(G_{F}^{-\delta}(K)G_{F}^{-\alpha}(K{+}L)-
    G_{\bar{F}}^{-\delta}(K)G_{\bar{F}}^{-\alpha}(K{+}L)\big)
    (g_{\delta\beta}(K{-}L)+g_{\beta\alpha}(2L{+}K)_{\delta}-g_{\alpha\delta}(2K{+}L)_{\beta}).
\end{align}

We can use the analogue of Eq.~\eqref{eq:feynman}
 for the anti-time-ordered propagator, i.e.
\begin{equation}
    G_{\bar{F}}(K)=-\frac{1}{2}\Big(G_{R}(K)+G_{A}(K)\Big)+G_{rr}(K)
    \label{eq:torderedprops}
\end{equation}
Using these definitions, we get 
\begin{align}
    \delta\mathcal{C}(l_{\perp})^\mathrm{virt}_{{\mathrm{Y}}}=&g^{4}C_{R}C_{A}\int\frac{d^{4}K}{(2\pi)^{4}}
    \int\frac{dl^{+}dl^-}{2\pi}\frac{i}{k^{-}-i\epsilon}G_{>}^{\beta-}(L)\delta(l^-)\nonumber
    \\&\hspace{-2.cm}\times\big(G_{rr}^{-\delta}(K)G_{R}^{-\alpha}(K+L)
    +G_{R}^{-\delta}(K)G_{rr}^{-\alpha}(K+L)\big)
    (g_{\delta\beta}(K{-}L)+g_{\beta\alpha}(2L{+}K)_{\delta}-g_{\alpha\delta}(2K{+}L)_{\beta}).
\end{align}
The virtual contribution is given by the sum of the two, that is
\begin{equation}
    \delta\mathcal{C}(l_{\perp})^\mathrm{virt} =  \delta\mathcal{C}(l_{\perp})^\mathrm{virt}_{\widehat{\mathrm{T}}}
    + \delta\mathcal{C}(l_{\perp})^\mathrm{virt}_{{\mathrm{Y}}}.
\end{equation}
In this case we do not enforce a semi-collinear scaling for the coft gluon, since, as we 
shall show, there is no double-logarithmic contribution. Furthermore,
the Wilson line integrations set $l^-=0$, thus making the interaction 
with the medium instantaneous. We rather assume $l^+\sim\lp\sim \kp\sim g T\ll k^+$, 
leading to 
\begin{align}
    \delta\mathcal{C}(l_{\perp})^\mathrm{virt}=&2g^{4}C_{R}C_{A}\int\frac{dl^+}{2\pi}
    \int\frac{dk^+d^2\kp}{(2\pi)^{3}}
 \frac{\frac{1}{2}+\nbe (k^{+})}{k^{+}}
    \frac{  G^{--}_{>}(l^+,\lp)}{k_{\perp}^{2}+m_{\infty\,g}^{2}}
    \nonumber \\
    &\times\bigg[\frac{\kp^2+\bkp\cdot\blp}{(\bkp+\blp)^{2}+m_{\infty\,g}^{2}}
    -\frac{\kp^2}{\kp^2+m_{\infty g}^2}\bigg].
    \label{partvirt_almostfinal}
\end{align}
The asymptotic masses at the denominator
have been included since we assumed $\kp\sim gT$. They arise from the $k^+\sim T$, 
$K^2\sim g^2T^2$ limit of the HTL propagators in App.~\ref{sec_conv}.
The $l^+$ integration can be carried out through the mapping to EQCD, as summarized around
Eq.~\eqref{treeAGZ}. It leads to 
\begin{align}
    \delta\mathcal{C}(l_{\perp})^\mathrm{virt}=&2g^{4}C_{R}C_{A}
    \int\frac{dk^+d^2\kp}{(2\pi)^{3}}
 \frac{\frac{1}{2}+\nbe (k^{+})}{k^{+}}
    \frac{Tm_{D}^{2}}{l_{\perp}^{2}(l_{\perp}^{2}+m_{D}^{2})}
    \frac{ 1}{k_{\perp}^{2}+m_{\infty\,g}^{2}}
    \nonumber \\
    &\times\bigg[\frac{\kp^2+\bkp\cdot\blp}{(\bkp+\blp)^{2}+m_{\infty\,g}^{2}}
    -\frac{\kp^2}{\kp^2+m_{\infty\,g}^2}\bigg].
    \label{partvirt_final}
\end{align}
We recognize that this expression is proportional to the leading-order, adjoint soft 
scattering kernel $\cc(\lp)^\mathrm{LO}_\mathrm{soft}$, Eq.~\eqref{losoft}. 
From this expression and its associated 
$\qhat$ contribution one can see that the $\vert\bkp+\blp\vert\gg\lp$ single-scattering 
region is free of double-logarithmic enhancements, as found in \cite{Liou:2013qya}.
Furthermore, for $\kp\sim\lp\sim gT$, which we have used for our derivation, the 
lifetime of the virtual coft fluctuation is long, so that multiple soft scatterings 
can occur within it. Hence, Eq.~\eqref{partvirt_final} should just be 
considered as the $N=1$ term in the opacity expansion of the virtual 
contribution to Eq.~\eqref{LMWCshiftfinal}. Indeed, we have checked that 
Eq.~\eqref{partvirt_final} can also be obtained from the virtual terms 
in Eq.~\eqref{eq:multiscat}, i.e. those proportional to $\exp(i\bxp\cdot\blp)$, further 
confirming the  soundness of Eqs.~\eqref{LMWschroshift} and \eqref{LMWCshiftfinal}.

\bibliographystyle{JHEP}
\bibliography{eloss.bib}
\end{document}